# Unidirectional Secure Information Transfer via RabbitMQ

**M.W.H. Maatkamp**

A minor thesis submitted in part fulfilment of the degree of M.Sc. in Forensic Computing and Cyber Crime Investigation with the supervision of Dr. Martin van Delden and Dr. Nhien An Le Khac.

School of Computer Science and Informatics

University College Dublin

16 December 2015

**Abstract**

Protecting computer systems handling possible sensitive information is of the utmost importance. Those systems are typically air-gapped with data diodes to assure that no information can physically flow back. Traditional computer protocols like HTTP or SOAP which are normally used to transport information between computers are typical bi-directional communication protocols and are thus unsuitable to be used over a data diode.

Currently the only commercially available protocols over a data diode sold by vendors are file-based protocols. Other protocols can be custom made but are expensive and proprietary. There are currently no open source solutions to stream data in a generic way over a data diode other than those file-based solutions.

Purpose of the dissertation is to research if open source technology can be used to mirror the contents of a messagebus over a data diode to get a cost effective security-proof and almost maintenance-free solution. and to further research if this technology can be used to transfer not only plain text data but also data sensitive by nature by using end-to-end encryption so that this information could even be admitted as evidence.

Method used to validate the research is a practical case study that shows how a sensor stream can send unencrypted and encrypted events over a data diode of arbitrary size via a message bus which are transparently and securely transferred and re-emitted internally without any kind of configuration management.

Results show that it is indeed possible to successfully mirror data from a Message Bus over a data diode and it is thus worthwhile to further invest in this technology.

**Table of Contents**

## 1. Introduction

The Dutch National Police collects information from all kinds of sources, from public (online) sources, commercial, other government departments in or outside the Netherlands and even from within the police force itself. The collected information flows from lower into higher zones of trust. The information itself can be classified as open data such as publicly accessible news and weather reports or even highly sensitive information such as evidence.

Protecting those computer systems handling sensitive information is of the utmost importance. Those systems are typically air-gapped with data diodes to assure that no information can physically flow back since these diodes have their hardware modified in such way that information can physically flow only from the lower side of the network into the higher side.[1] These networks are commonly called "black" for the outside components and "red" for their internal part.

Currently the only protocols over a data diode sold by vendors are file-based protocols like (S)FTP, Mail and and Samba[2][3][4][5]. While other protocols can be custom made by vendors but are (very) expensive, proprietary and the intellectual property rights of those custom made software could even stay at the vendor of the application. Some vendors even support one-way TCP connections over a data diode. TCP protocols like HTTP or SOAP which are normally used to transport information between computers are bidirectional protocols which makes this solution unsuitable to directly interface such protocols.

A literature study however showed that there are currently no open source solutions available to stream data coming from a message bus over a data diode. Even on github, the place where open source projects are hosted no projects could be found which provide this functionality.

RabbitMQ is an open Open source and commercially supported messaging broker and and uses Advanced Message Queuing Protocol (AMQP) as a protocol which makes it interchangeable with other Message Bus platforms like ActiveMQ.

This paper is the evaluation of RabbitMQ as a message bus to be used transparently over a data diode to get a cost effective and almost maintenance-free solution and to further research if this technology can be used to transfer not only plain text data but also data sensitive in nature by using end-to-end encryption so that this information could even be admitted as evidence.

During this research we found and implemented such a solution on how to transport messages in a transparent way over data diode even when encryption is used making the diode functionality literally transparent.

Results show that it is indeed possible to mirror data over a data diode into a secondary message bus and it is thus worthwhile to further invest in this technology.

## 2. Literature Survey

The reason for starting this research is that there were no other protocol solutions being offered by commercial vendors of data diodes other than SAMBA, FTP, FTP/S, SMTP (Email), TCP and UDP. While these protocols are normally sufficient for file transfer none of the options provide a streaming solutions other than TCP.

The second place to look for the possible existence of research for such a protocol is google with the focus on searching for the terms “data diode” together with the terms “esb”, “message bus” or “RabbitMQ” which also yielded no practical solutions.

Google Scholar however turned up a document called “Demonstration of a Cross Security Domain Service Management Capability for Federated Missions “[7] which talks about using ESB’s in a closed military environment but does not disclose code or the exact details how their custom solution works but ends with this interesting conclusion: “*A key lesson learnt from this experimentation process is that the service bus messaging pattern is a critical enabler in implementing a distributed event and configuration management system. It can provide persistence, throttling and caching and a common internal interface to the cross domain gateway. This way multiple different ELMs can connect to the IOSS while the cross-domain solution remains transparent and possibly interchangeable or upgradeable in the future.*” which is exactly why this research was started in the first place.

While trying to find open source alternatives github.com was also queried but the query "data diode" returns only one result which is a project how to turn two Cisco switches into a data diode.

## 3. Problem Statement

Normally when two networks are being connected a firewall is put between them. A firewall is a network security system, either hardware- or software-based, that controls incoming and outgoing network traffic based on a set of rules. Because these rules can be added, changed or deleted by an administrator (or by a hacker or malicious piece of software) firewalls could possibly contain errors in their configuration and thus need to be constantly monitored.

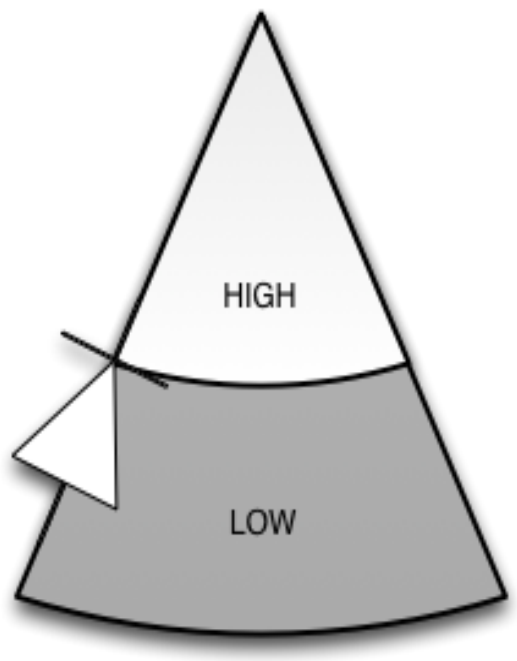

A data diode on the other hand is a device where data can physically flow from one side to the other because hardware modifications have been made to make it physically impossible for data to flow back [1]. These hardware modifications can be done on many types of media like RS-232 and even optical fiber where the idea is that the higher side can only listen and the lower side can only send information.

In the case of an optical fiber solution, there is no fibre optic link between the transmit port of the transceiver on the receiving side of the data diode and the receiving port of the transceiver on the transmitting side of the data diode. Visual inspection of the transceivers should easily show that they have not been tampered or replaced with a component that can both transmit and receive.

A protocol like **Transmission Control Protocol (**TCP) is unsuitable to be used because TCP is a bidirectional connection-oriented protocol. A protocol more suitable to be used over a data diode is **User Datagram Protocol (**UDP) because UDP is specifically designed as a unidirectional communication protocol.. Because UDP has notion of state it provides no mechanism to guarantee that packets arrive in the same order as they were being sent but also provide no guarantee that packets arrive at all. The only way for the receiving side to know that information is missing is to include a sequence number in the payload which is also needed for data to be admitted as evidence.[7]

RabbitMQ is an open Open source and commercially supported messaging broker and and uses Advanced Message Queuing Protocol (AMQP) as an Open Standard for Messaging Middleware which makes it interchangeable with other Message Bus platforms like ActiveMQ. It supports clustering and is thus fault-tolerant and uses flexible routing. RabbitMQ is officially supported on a number of operating systems and several languages with the most prominent being java which has excellent support by the way of Spring Boot/Spring AMQP which we will use in this dissertation.

RabbitMQ acts like a mailbox where messages can be delivered to multiple recipients based on headers provided with the message. A producer is a process which sends messages and a consumer is a process which waits for messages. An exchange waits for messages and sends them to one or more queues and queues are the buffers which stores messages which are not yet consumed. In an ideal setup all the queues are consumed by consumers and are thus empty.

Within RabbitMQ:

- the core is a publish/subscribe mechanism
  For a creating process when a message is published it is being delivered as fast as possible to the receiving processes without any lag, which would be introduced by polling. For the receiving side it means that an individual receiving process can subscribe itself on events and the message bus handles the publishing of newly created messages to all receiving processes by copying and delivering the messages in parallel.

- supports extra header information
  Just like email extra information can be added to a message like routing keys or headers containing mime-types. Receiving processes can filter on those or use it to add extra meta-information with the contents of the message when added in a database for example.

- support message filtering
  receiving processes can subscribe itself to get all messages but there is also the option to get a subset of messages. These are filtered on routing keys or specific headers.

- support logging and monitoring
  Since the core of a message bus are messages, these busses also expose information like message rates and queue sizes (an incrementing queue size indicating a faulty receiving process) can all

- support users and groups
  A receiving process gets only the data from exchanges a process has being given rights to by the manager of the message bus and thus it can not do a directory scan like in a file-based solution it would be possible for a receiving process.

- support transactions
  since the core of a message bus are messages, these messages are kept in memory until the message bus gets an acknowledgment that the message is being successfully transferred to the receiving process after which the copy is deleted. Could a message not being delivered because of an error, the message is retransmitted to another process or is being put back in the queue.

- support queueing
  could a message not being delivered because of a timeout or error because the receiving process is no longer there for example, messages are being held in a queue for as long as
    - a message Time To Live (TTL) signals it has not expired and
    - there is enough memory in the message bus host itself

When the receiving process reattaches itself to the bus, all queued messages are being delivered to the process. The sending process could in this scenario deliver its messages even when the receiving process was no longer there and the receiving process still gets all the messages even when the process was momentarily inaccessible due to an error or an upgrade of the receiving process itself.

Messages themself can be bigger than UDP allows and have to be split into smaller segments.

**Message splitter**

Messages which are transferred over RabbitMQ can be bigger than the standard UDP packet size of 1500 bytes. There has to be a system in place which splits the messages up into smaller segments to an adjustable size and adds a header with all the bookkeeping information like original message size and preferably a checksum to validate that the original message is transferred correctly. On the receiving side the message should be reconstructed and validated by its checksum.

**Redundancy**

Since packet delivery in UDP is not guaranteed, data should be sent redundantly to cope with lost packets. The redundancy factor should be adjustable and a higher number means more data is send redundantly which makes the system more fault tolerant but at the cost of transfer speed.

**Cleanup**

When packets arrive some get lost. The system has to do bookkeeping to reconstruct the messages but when too many packets are missing it cannot reconstruct the original message and would wait indefinitely for those packets to arrive. Periodically it should be checked whether a message gets new packets but the message should be discarded whenever a certain amount of time has passed and no new data is being added to the message. It should clean up and print a line in the logfile that a message was discarded. Should this happen, the throttle could be set higher and data can be send even more redundantly.

**Encryption**

Connections to a message bus can be forced to use SSL and user-permissions like read and write permissions can be set on individual exchanges for specific users but all messages are up till this point still readable by anybody with administrator privileges.

For information to be admitted as evidence authenticity and integrity are needed [8][9]. We want to verify that encryption of individual messages can also used in this environment and should this research be successful the encryption itself will most likely be replaced by an infrastructure already present, but for now the encryption we are going to use will be a small encryption framework which proves the point of encryption, integrity and authenticity but is not industrial hardened against any possible attack

Regular symmetric key encryption with a shared key like AES does not prove the integrity nor authenticity of the message because nothing prevents an attacker from generating a random message which the receiver will then decrypt and accept.[12] And when a symmetric key is compromised all previous messages become also readable. Therefor a public/private keypair for the server and for the sensor have to be generated and used to encrypt a new symmetric key for each message to provide:

- *Integrity* - A hashing algorithm is used to calculate a hash of the message and when received a difference between a calculated hash and the hash of the message indicates whether a message has been modified.

- *Authentication* - To prove that the communication does indeed originate from an identified party (sender authenticity) the sending party signs the hash with the private key so that the receiving party can validate with the public key (preferably being given out-of-band that the message indeed did come from the sender.

- *Non-repudiation* - Non-repudiation means that after a message is signed and sent one cannot deny having signed the original message because the message was signed with the private key of the sender which hopefully nobody else knows.

The sending process should encrypt the data with a new symmetric key for each message, calculates a hash over the encrypted contents and signs the hash with the private key of the sensor. The symmetric key is then encrypted with the public key of the inside process and the message is transferred over the data diode. This scheme is called 'Encrypt-then-MAC and means that ciphertext is generated by encrypting the plaintext and then appending a MAC of the encrypted plaintext which is provably secure [12].

In this research we want to prove that a smart device (a device which is e capable of doing their own encryption) can encrypt its contents and add enough header information for the receiving process to determine if the message is to be decrypted and can validate if the sensor did indeed send its information by checking the signed hash of the message. The device should therefore contain not only its public/private keypair but also the public key of the inside process. The receiving process on the inside validates the message and when valid decrypts and send the information the exchange where it originally came from on the outside or give an error-message when validation did not succeed.

We also want to prove that a non-smart device (which is a device which does not have the computing power to do encryption on its own) can send to its information to the a message bus to the "encrypt" exchange where it gets encrypted and send over a data diode by a separate process. Incoming messages are being encrypted, signed and send over a data diode where a decrypting process decrypts the messages and will validate and send them to the exchange where they originally came from.

**Jumbo frames**

Jumbo frames can be enabled to send bigger packets over the data diode. Normally a network would declare the maximum size of an UDP packet to be 1500 bytes, but some drivers have the option to enable jumbo frames which can be 9k, 16k or even 64k but those are called super jumbo frames and are very rare.

**Throttling**

We want to be capable of throttling the speed at which UDP messages are being sent over a data diode since packets get missing because the receiving side cannot keep up with packets arriving at a certain speed.. The software should contain a property which indicates the time to wait before sending another packet as not to overflow the receiving buffer which would then drop the packets.

## 4. Adopted Approach

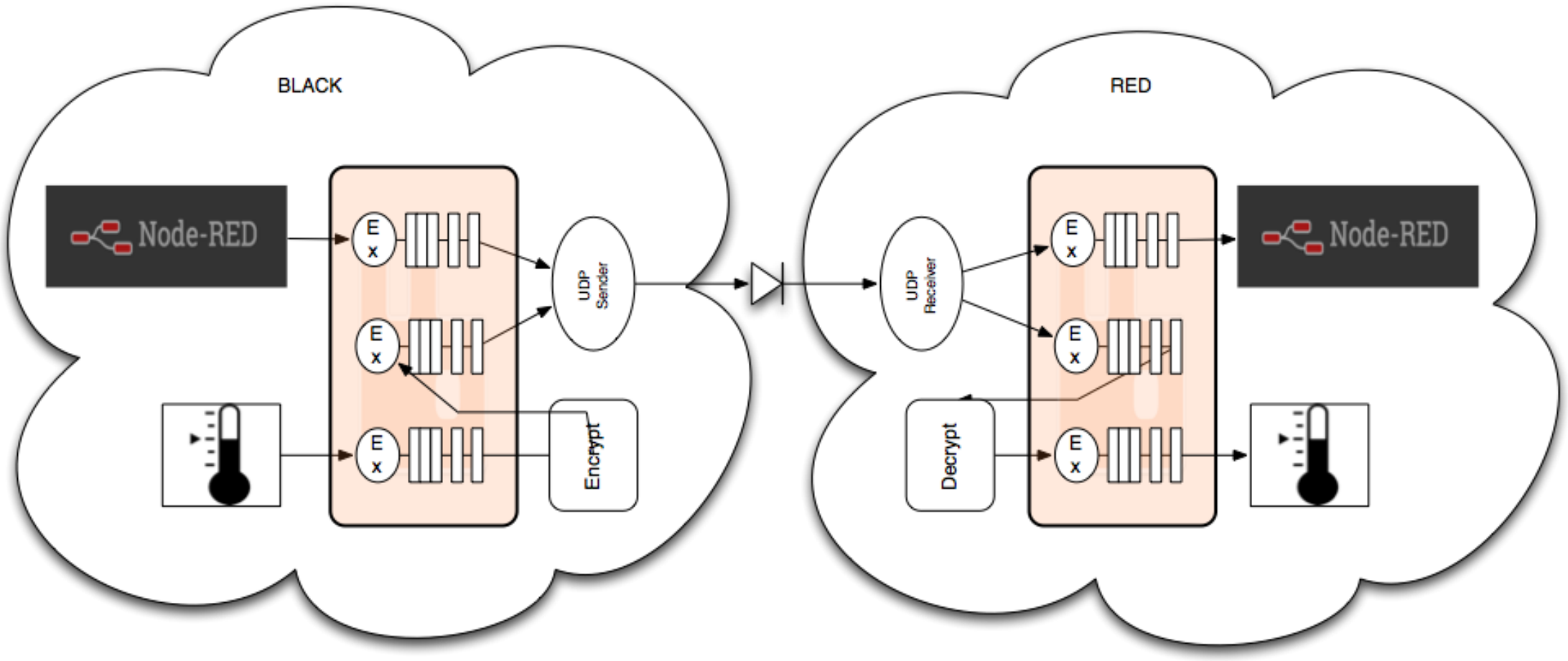

Fig 1: schematic overview

Above is the schematic overview of the infrastructure what we want to implement. We want to research if RabbitMQ messages can be send over a diode and do that with a fictitious temperature sensor which sends periodically a temperature sensor event over to the red side. The sensor can send its data in plain text messages to an exchange which are mirrored over the data diode and also by sending its data to the “encrypted”-exchange where the data is encrypted before it send over to the red side.

The idea is that this is a real-world example of a sensor which we can use to develop and test the serialisation and encryption and decryption of such messages.

This dissertation describes what has been developed to get RabbitMQ messages transparently over a data diode:

| **Application** | **Description** |
|---|---|
| Automatic Queue creation & Listeners | On the red side runs an application which checks periodically if new exchanges are created and attaches a listener to a queue bound to the exchange to get the messages |
| Encryption | All messages routed to the exchange “encrypt” are being encrypted with the key of the receiving side and signed with the key of the sender |
| Cut application | The cutter splits those messages into smaller messages and appends the exchange and type of exchange the message came from and adds a header to it containing the UUID and index and calculates a checksum |
| UDP Sender with Rate Limiter | The UDP Sender transfers messages over UDP. When the sender sends UDP packets too fast over the network, packets get lost. A rate limiter limits how many packets per seconds are being transferred. |
| UDP Receiver | A dedicated c program has been developed to transfer UDP messages into RabbitMQ. At the measured speeds of over 12.000 messages/sec the garbage collection of an interpreted language like Java and NodeJS cannot keep up, but the C implementation proved to be stable. |
| Merge application | Reconstructs the message and calculates the checksum. If the checksum matches the checksum in the message, the exchange is created and the message is delivered to the exchange. |
| Decryption | When messages are reconstructed and put in the exchange “encrypted” those messages are decrypted with the private key of the receiver and validated with the public key of the sender and if the validation succeeds the message is decrypted message is being transferred to the original exchange. |

**WORKFLOW**

This is the flow of a sensor which sends its data over in plan text.

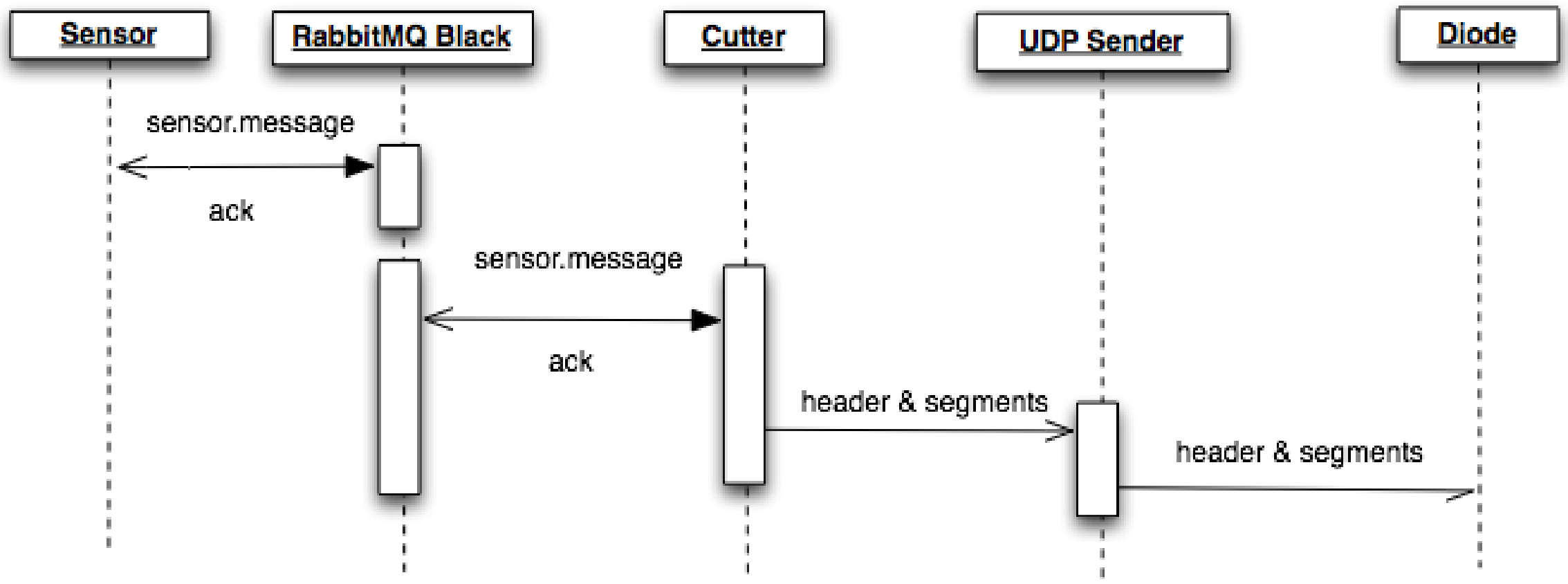

Fig 2: Sending a message

When a sensor is delivering a message to the black side it sends its messages to the exchange. A Message Listener is automatically attached to this stream and wraps the message in an Exchange Message adding extra meta-information of the exchange so that the receiving side has enough information to reconstruct the exchange should it not exists and sends the message to the UDP Sender. This takes the message, serialised it into a byte stream and sends the stream over the diode.

This is the receiving side of a sensor event in plain text.

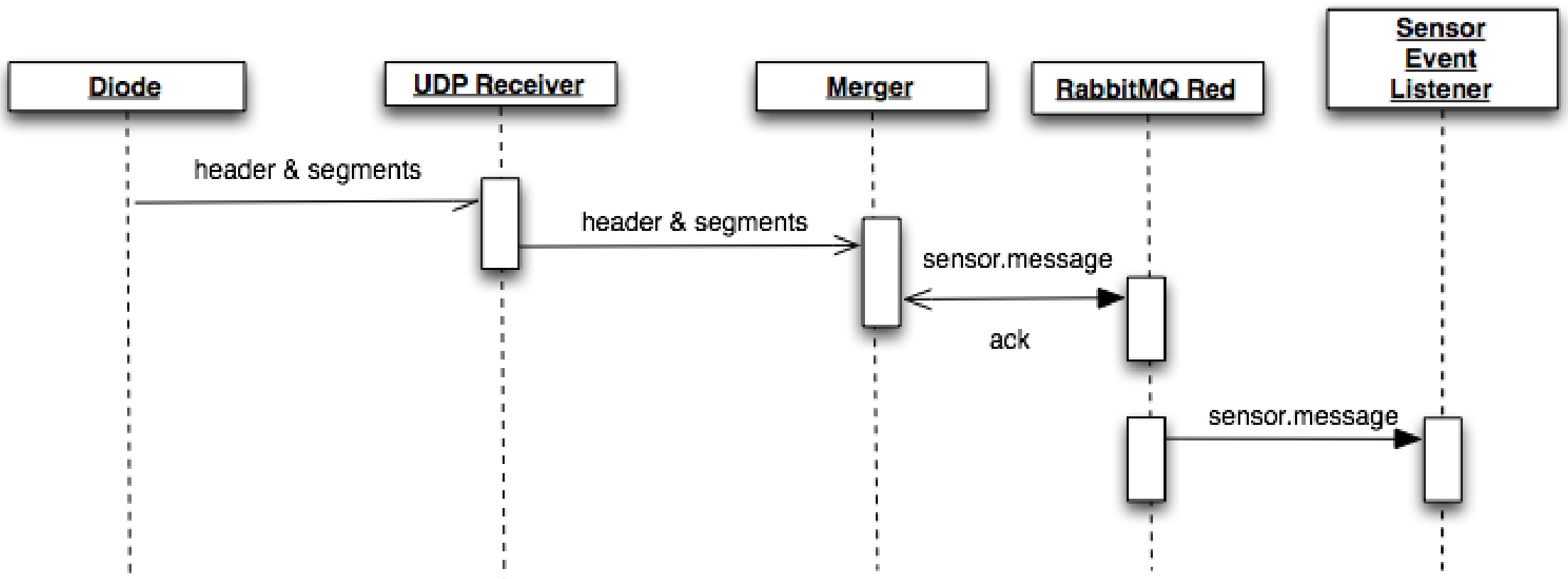

Fig 3: receiving sensor message

On the receiving side the UDP Receiver reconstructs the exchange message from the bytestream and hands it over to the RabbitMQ Service. This service reconstructs the exchange should it not exist and sends the message to the exchange. In the case of a sensor event a listener picks up those messages and prints the contents of a sensor message in the console.

**Messages**

The core of AMQP is org.springframework.amqp.core.Message which contains a field “body” with the content and a field org.springframework.amqp.core.MessageProperties containing fields used for delivering a message such as a routing key and other message properties like mime-type.

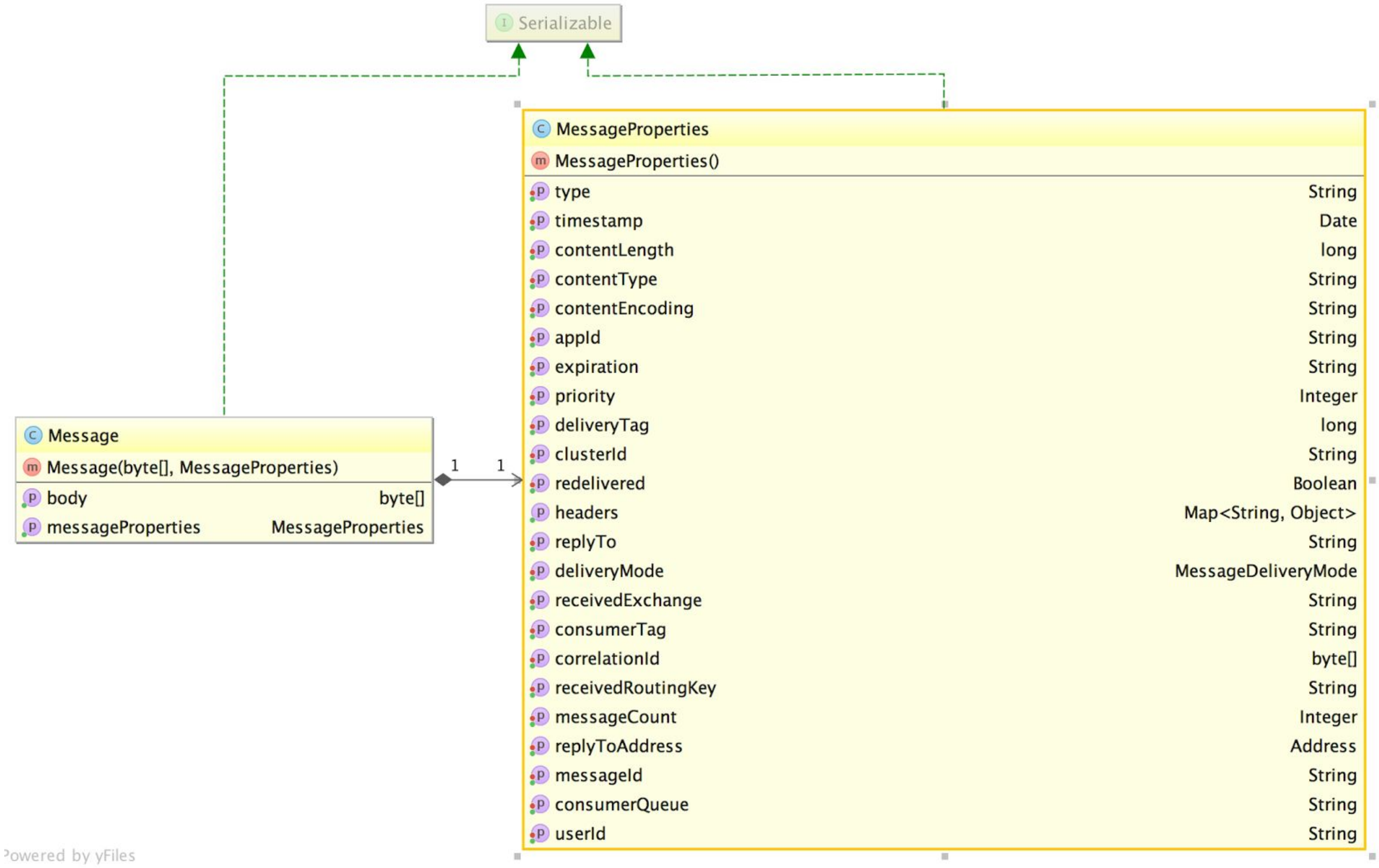

Fig 4: Message and MessageProperties.java

A message is like an email message: it contains a body, where it should be send to (in this case an Exchange, which is something like a mailbox but is of a certain type) and something which is called a “routing key” which is an arbitrary field on which a receiving process can make the decision that the message might be of interest.

The idea is to mirror this exact message on the inside RabbitMQ. The problem is that Message contains a field “receivedExchange” which contains only the name of the exchange where it came from, but does not contain the actual type of an exchange. While this is enough information when the exchange is already created (which is the typical use-case because an exchange has to be created before a Message can be send to it) it is not enough to reconstruct an exchange on the red side.

Since org.springframework.amqp.core.Exchange does not inherit from Serializable.java, is is impossoble to send it without modification over to the red side Another class is needed which contains a copy of the contents of a instantiated amqp.Exchange.java. This is org.datadiode.model.message.ExchangeMessage where the contents of amqp.Exchange.java is serialized with XStream into a field “exchangeData” of type String which contains in JSON-format all the contents of all the fields of an amqp.Exchange. With this information the red side is now capable of reconstructing the Exchange with all of its properties and send the contents of the Message to the newly created Exchange over on the red side.

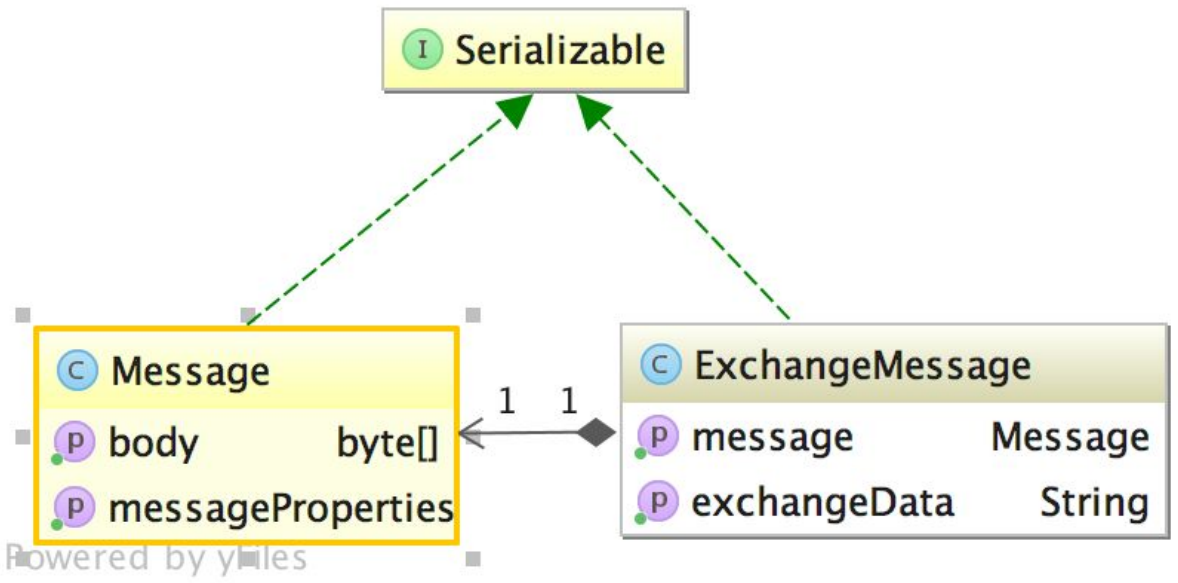

File X: EchangeMessage.java

The contents of exchangeData looks like the following figure where type=FanoutExchange, name=”sensor”, durable=”true” and autoDelete=”false” are all parameters necessary when creating an Exchange:

exchangeData.json Raw

```json
{
   "org.springframework.amqp.core.FanoutExchange":{
      "shouldDeclare":true,
      "declaringAdmins":[
         {
            "@class":"list"
         }
      ],
      "name":"sensor",
      "durable":true,
      "autoDelete":false,
      "arguments":[
         {
            "@class":"linked-hash-map"
         }
      ]
   }
}
```

Fig 5: exchangeData.json

The application looks up if an exchange with that name exists, if not creates one and sends the message to the newly created Exchange.

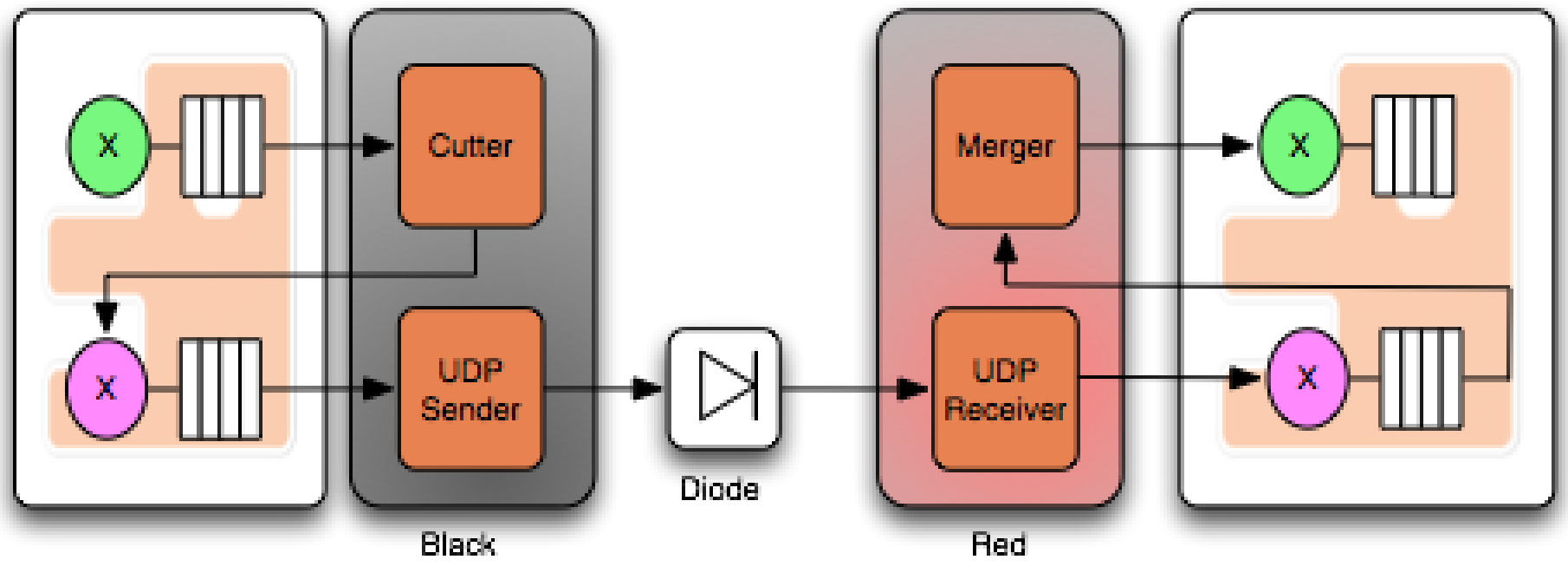

Fig 6: messages mirrored over a data diode

The application at startup periodically creates new queues with the same name as any (newly) created exchange but it an extra ".dd" appended to hint that these queues are being used for sending data over the diode. It attaches listeners to all these queues and when a message from a queue is received the listener adds enough information necessary for reconstruction of the exchange and message on the red side and sends this as a UDP packet over a diode.

Also on the receiving side there is a receiving process which takes the information from UDP, transforms the information into a Message, recreates the Exchange if it is not there yet and sends the message to the Exchange thus effectively mirroring the RabbitMQ over a diode.

**Message splitter**

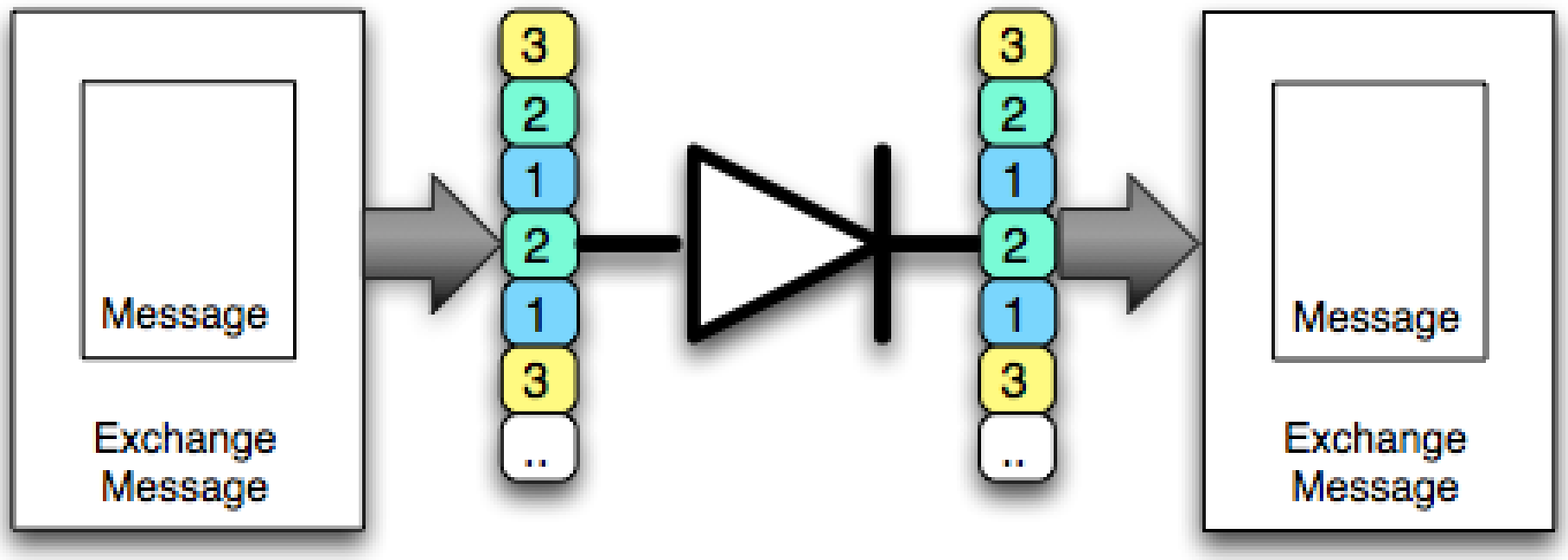

Fig 7: Message is split in header and segments

Messages can get bigger than the standard UDP message size of 1500 bytes or 8K if jumbo frames are enabled. To transfer those messages they are split up into smaller chunks of a predefined length before they are transferred over UDP. This is done in the Cutter-application which splits messages into a specified amount of bytes, adds a header and index and adds the data redundantly and shuffles the redundant data to mitigate packet loss. The length of an individual is determined by the following property:

```
# set this (MTU - 29 bytes header)
application.datadiode.cutter.size = 8163
```

The receiver side then removes duplicates and reconstructs the packets by ordering the segments on the received index  and checks the checksum of the reconstructed message with the checksum in the header. If those match the message is then resend into the red side to the specified exchange with the original message size.

**Segment**

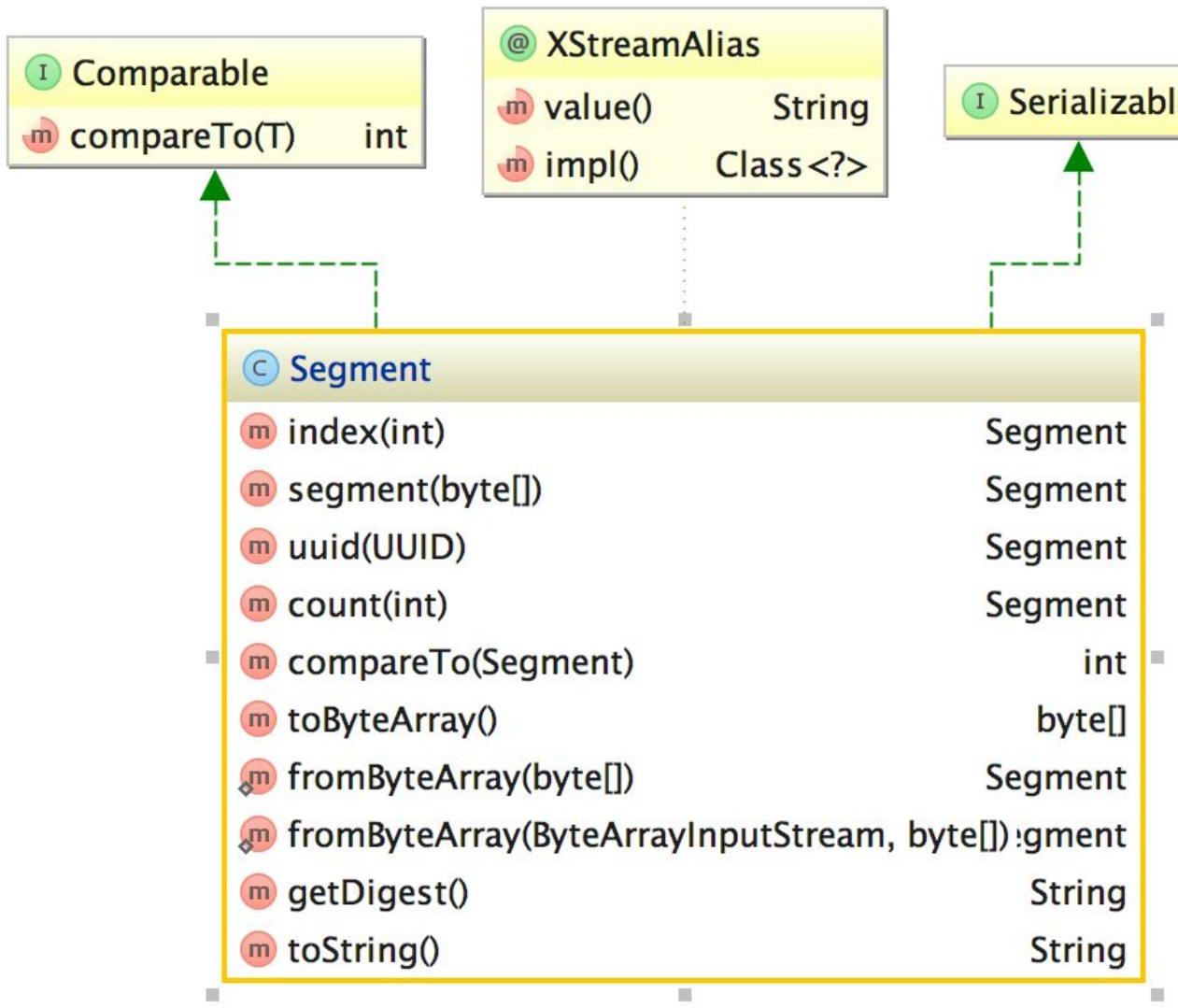

fig 8: Segment.java

A segment can be identified by the first byte and contains an UUID which is the unique identifier for each transmitted message. The backend adds each segment to the list of other segments for the same UUID and when the count of those received segments equals the number of the count variable in the message the message can be reconstructed. Because packets can arrive out-of-order an “index” in the segment determines the packet order. The backend then reorders all the segments on the value of the index and the message is ten reconstructed and validated by the checksum. The segment length is used to determine the actual size of the data in the packet. So each packet contains an extra 29 bytes header information.

| INDEX | NAME | TYPE | BYTES |
|---|---|---|---|
| 0 | type | BYTE | 1 |
| 1 | UUID | LONG * 2 | 16 |
| 17 | count | INT | 4 |
| 21 | index | INT | 4 |
| 25 | segment.length | INT | 4 |
| **TOTAL** | | | **29** |

**Redundancy**

Should packets disappear due to a network error the data can still be reconstructed because packets can be added multiple times. This is done with the redundancy factor which is 2 in this example where every packet is inserted twice in the stream. The property to adjust this value is

```
application.stream.cutter.redundancyFactor=2
```

**Shuffling**

When there is packet loss packets disappear. To minimise the impact of packet loss redundantly inserted segments are shuffled randomly to minimise the impact of packet loss:

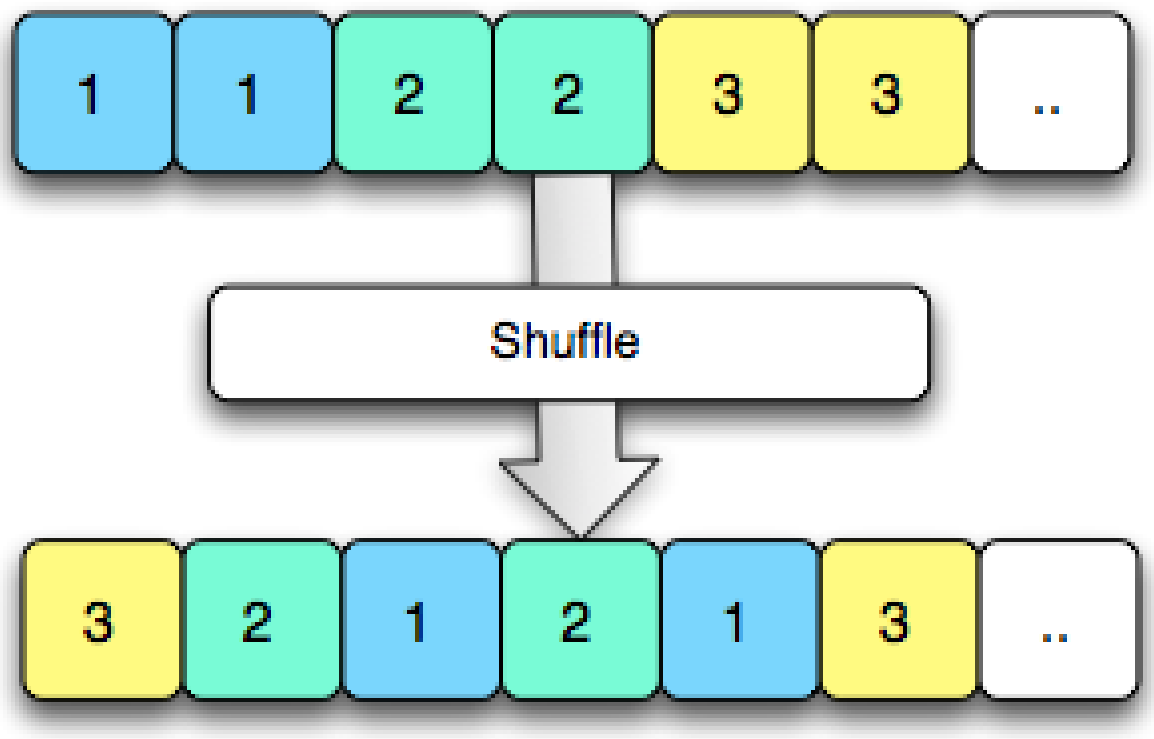

**UDP Sender**

While testing the application it showed that the java implementation could not keep of with message rates above 12.000 messages/sec of 8K each and caused packet loss. Another implementation in NodeJS had the same problem which also caused packet loss at the receiving side. A dedicated C program was developed which publishes UDP messages to RabbitMQ which solved the problem and proved to be stable at over 17.500 messages/sec with no packet loss.

**Reconstruction**

In the code the deduplication and reordering is done automatically by inserting the messages in a Treeset:

```
Map&lt;SegmentHeader, TreeSet&lt;Segment&gt;&gt; uMessages = new ConcurrentHashMap();
```

Treeset has 2 unique features: it contains only unique items and those items are ordered which is exactly what we want: we want deduplication and reordering. By inserting duplicate items in the TreeSet those duplicate items are automatically removed and the set is kept in order.

Whenever there is packet loss this is mitigated because packets are inserted multiple times in the stream so that the original message can still be reconstructed even when some segments are missing:

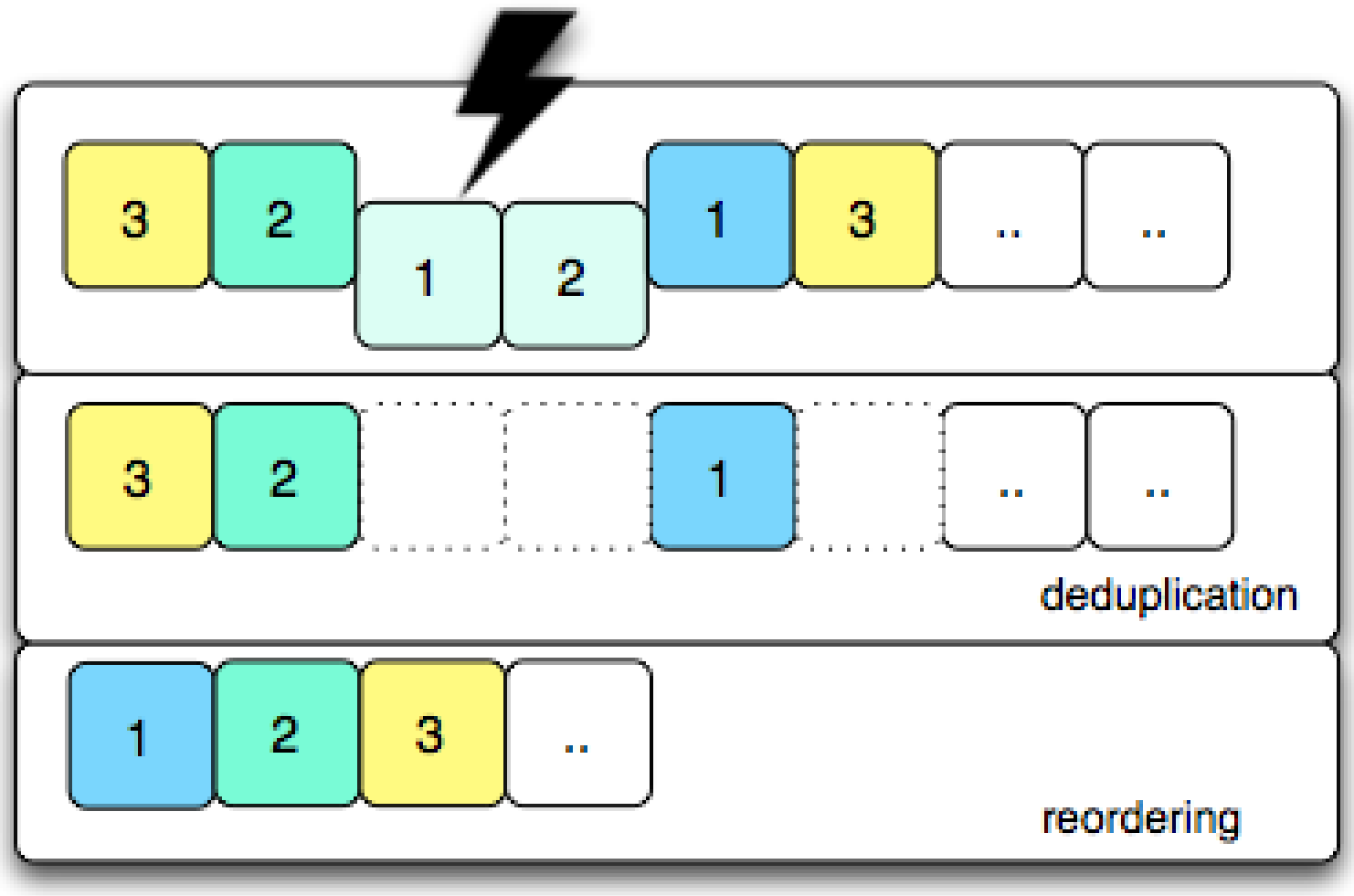

Fig 9: Packet reorder and packet loss

With all these measures in place this is effectively what we now have created: a two RabbitMQ instances where the contents on the outside RabbitMQ is perfectly mirrored on a RabbitMQ on the inside:

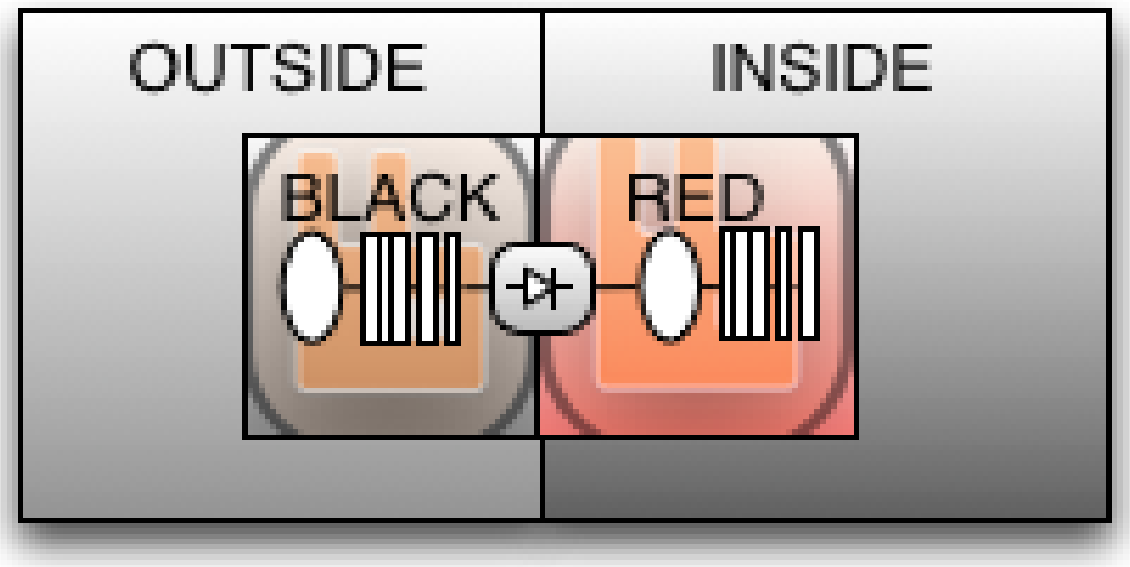

Fig 10: RabbitMQ mirrored

**DMZ**

Now that this setup works, a more typical scenario would be a mirror the data first in a DMZ and let somebody from the OPS department decide if the data stream should be mirrored into the inside of a network by using so called "shovels" which by definition of RabbitMQ is a plugin and its function is "*to reliably and continually take messages from a queue (a source) in one broker and publish them to exchanges in another broker (a destination).*"

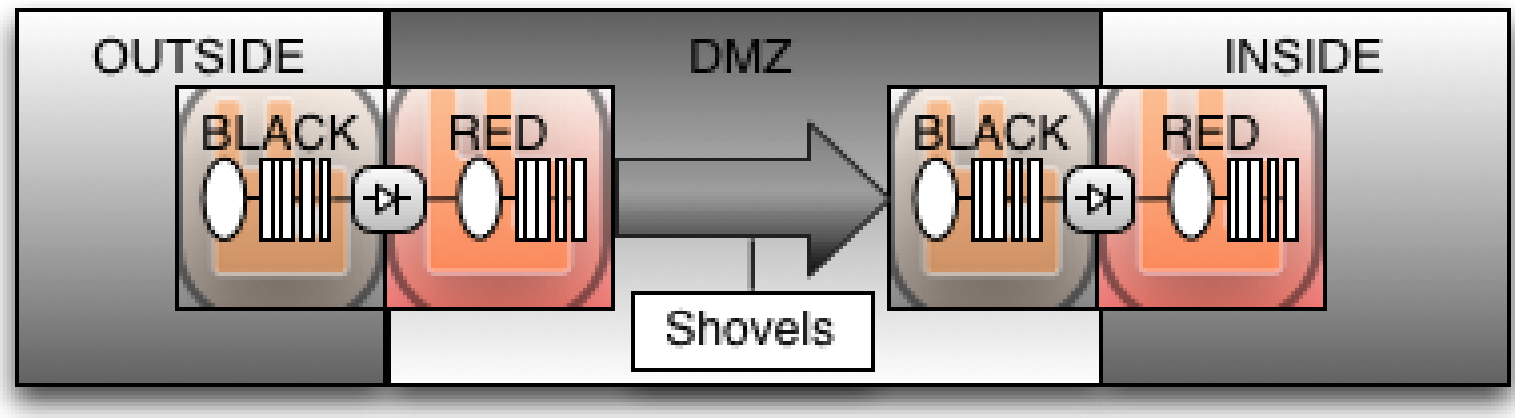

Fig 11: RabbitMQ in a DMZ

The idea behind this rationale of a DMZ is that all not all messages are to be transported into the inside Only messages coming from known exchanges can be brought to the inside but **only** when somebody in person adds a shovel from the outside to the inside in the management interface on the black side on the inside mirror. Should messages being added by an unknown party, those messages will thus stay at the red RabbitMQ of the outside mirror for further inspection

**Performance property**

The problem with UDP is that it does not guarantee delivery of packets. Normally packets are dropped when the receiving side is not capable of keeping up with the arrival speed and the sending process needs some way to throttle the message rate. A rate limiter has been implemented and the rate can be adjusted with the following property to make it faster or more reliable:

```
application.datadiode.udp.external.rate = 14500
```

**ROUTING**

All headers from a message are being transferred into the inside. This means that the routing key is also being transferred into the inside. A routing key can contain everything but contains an identifier where, at the inside, a message can be routed on without decrypting the message content itself.

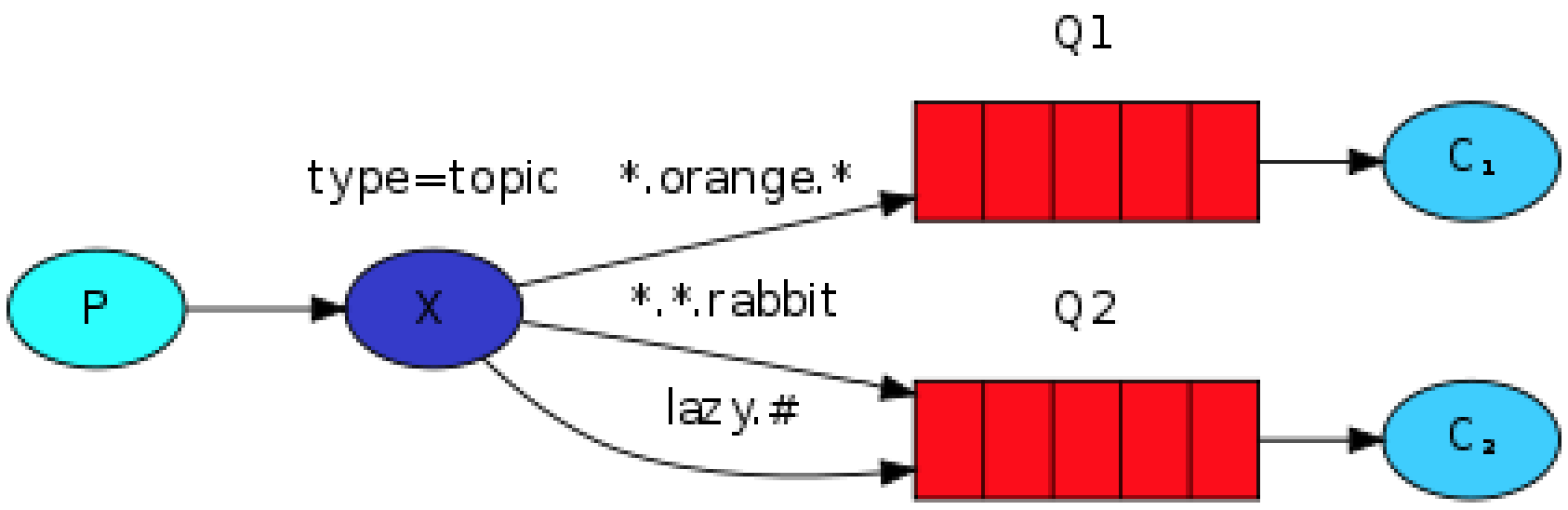

Fig 12: routing keys

**ENCRYPTION**

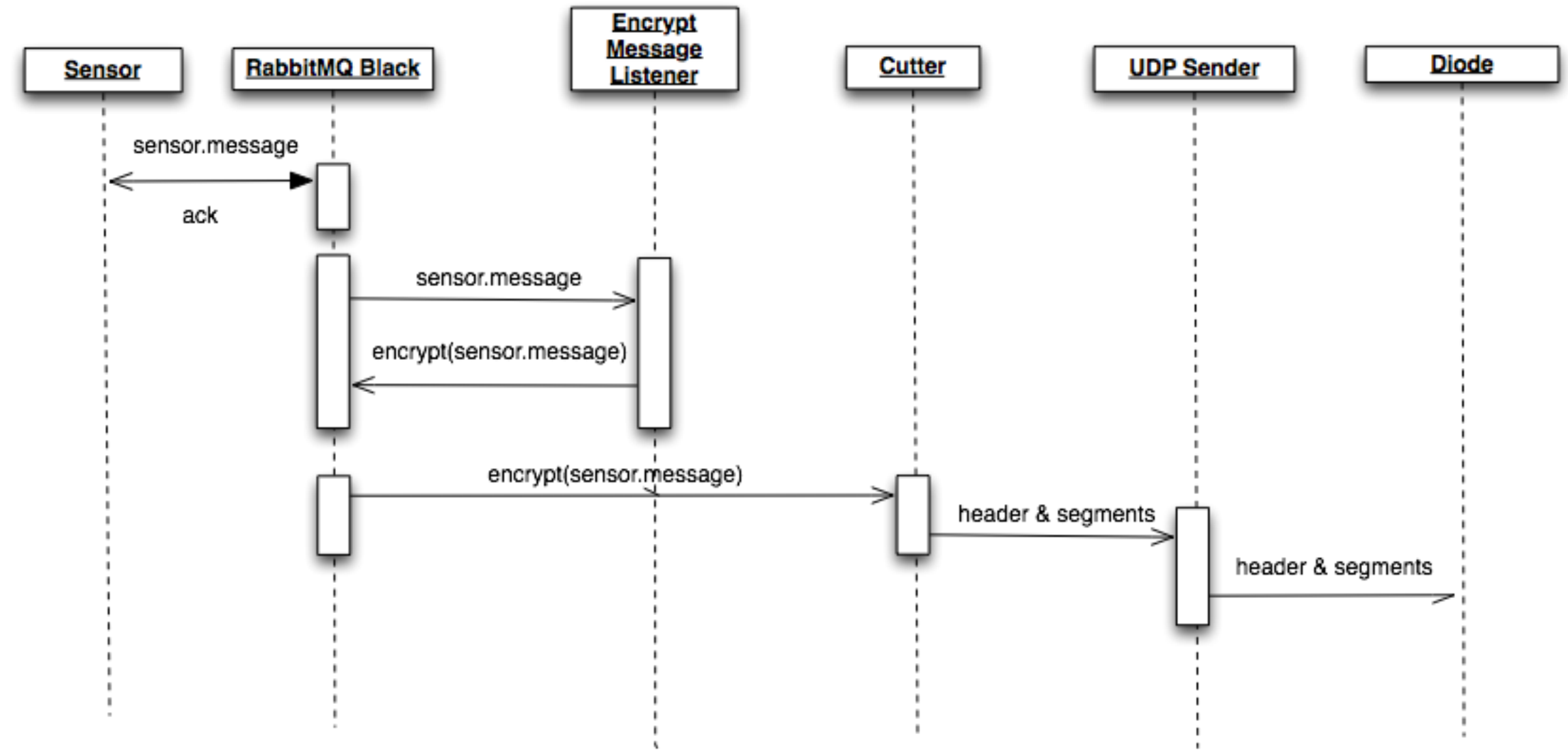

Fig 13: sending encrypted message

Here an unencrypted sensor sends its data to the RabbitMQ on the black side to the "sensor"-exchange. A shovel shovels the message over the the the "encrypt"-exchange where a listener "EncryptMessageListener" gets the message and encrypts and signs the message and sends the resulting message back to the RabbitMQ to the "encrypted"-exchange. There a generic MessageListener picks up the message and encapsulated the message with a exchange-message containing the original exchange and sends the result over UDP via the UDP-Sender over the data diode to the red side.

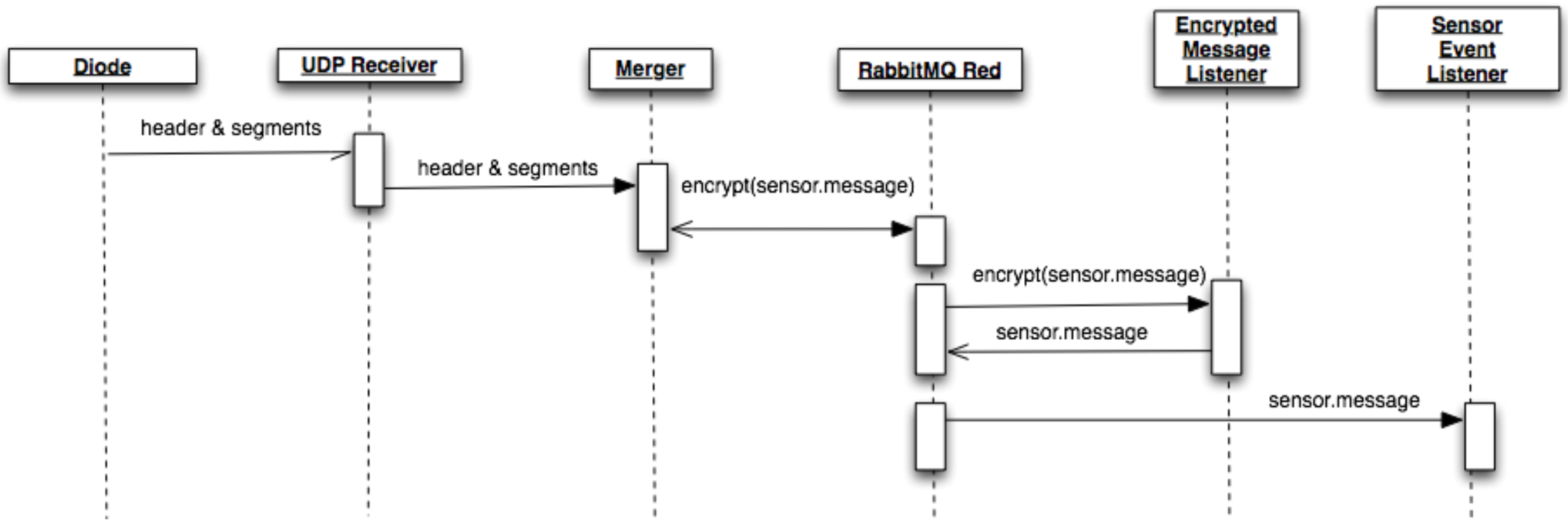

Fig 14: receiving encrypted message

There a UDP-Receiver receives the message and reconstructs where the message came from: the “encrypted”-exchange. It sends the encrypted message to this exchange where the EncryptedMessageListener listens for new messages and gets the  the encrypted message which is validated and decrypted and is then send to  the original “sensor”-exchange

Which encryption method is used and the key sizes are specified in application.properties.

| | |
|---|---|
| black | src/main/resources/application.properties |
| red | src/main/resources/application.properties |

application.properties Raw

```
application.datadiode.red.cipher.provider = BC
application.datadiode.red.cipher.signature = SHA256withRSA

application.datadiode.red.cipher.asymmetrical.algorithm = RSA
application.datadiode.red.cipher.asymmetrical.cipher = RSA/NONE/PKCS1Padding
application.datadiode.red.cipher.asymmetrical.keysize = 2048

application.datadiode.red.cipher.symmetrical.algorithm = AES
application.datadiode.red.cipher.symmetrical.cipher = AES/EAX/NoPadding
application.datadiode.red.cipher.symmetrical.keysize = 256
```

The first property is the provider, in this case BouncyCastle which is an alternative to the standard cryptographic library called Java Java Cryptographic Extension (JCE) which contains many more cipher suites and algorithms.

The second property is the way messages are signed. Used here is “SHA256WithRSA”  which means that a message is hashed with SHA256 and signed with a RSA private key.

The message itself is encrypted with AES-256 and uses the “AES/EAX/PKCS7Padding” where EAX is a *“mode of operation for cryptographic block ciphers. It is an Authenticated Encryption with Associated Data (AEAD) algorithm designed to simultaneously provide both authentication and privacy of the message (authenticated encryption)“* [13] and uses padding for the last block.

The key used to encrypt the message is then encrypted with the public key of the server with RSA using 2048 bits and uses padding for the last block. Should these parameters be not sufficient just modify application.properties and start the application again.

When the application starts it will start a org.datadiode.black.listener.EncryptMessageListener which is listening for messages being send to the exchange “encrypt”. It will encrypt these messages with the public key of the red side, signs the hash with the private key of the black side and emits the encrypted message to an exchange “encrypted”. The exchange “encrypt” is specifically not mirrored but the exchange but the exchange “encrypted” is.

On the receiving side a listener will decrypt the contents of the messages coming from the mirrored exchange "encrypted" with the private key from the red side and validates that the message really came from the black side with the decryption of the hash with the public key from the black side and if valid sends the message to the exchange it came from on the black side.

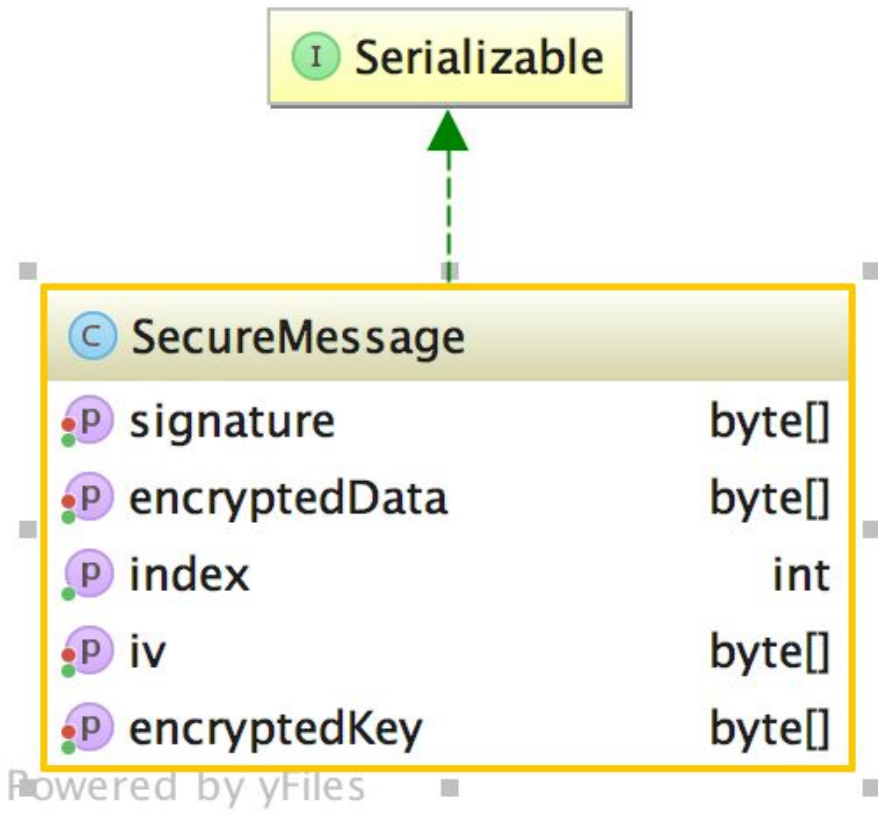

Fig 16: SecureMessage.java

This solution has been made so generic that it can be used to encrypt any stream of data by shoveling data from any queue into the "encrypt"-exchange.

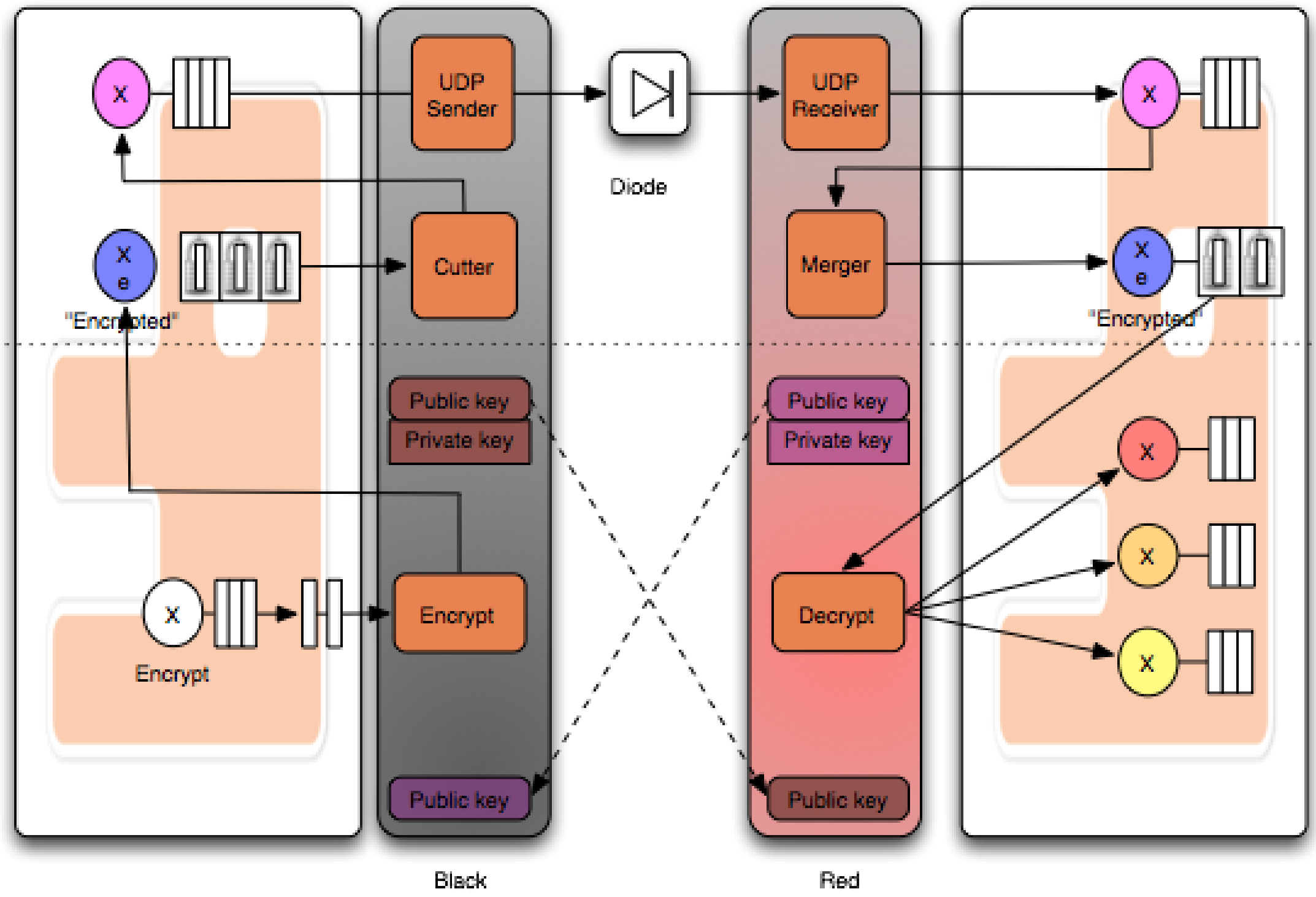

Fig 17: Unencrypted and Encrypted stream are both mirrored

Messages from outside can be send to an exchange at the black side. The creation of new exchanges is being picked up a new instance on each of GenericMessageListener is attached which starts to broadcasting messages over the data diode. On the receiving side the UDP receiver transforms an UDP message back to the original message and is being transmitted to the exchange it originally came from. It also watches an index in the message to detect missing packets by comparing the index with a counter containing the number from the previous message. Should the value of index not be the the same as the previous index plus one, a warning is being displayed in the log of the application.

When encryption is used a message is thus first encrypted into an EncryptedMessage which in turn is added into a ExchangeMessage so that it can be routed and that exchange message is split into smaller chunks (a header and some segments) which are the units being transferred over a diode via UDP.

The receiving side then reconstructs the ExchangeMessage and validates the checksum after which it sends the EncryptedMessage to the EncryptExchange. This is being picked up by the listener, decrypted and validated and resend to the original exchange.

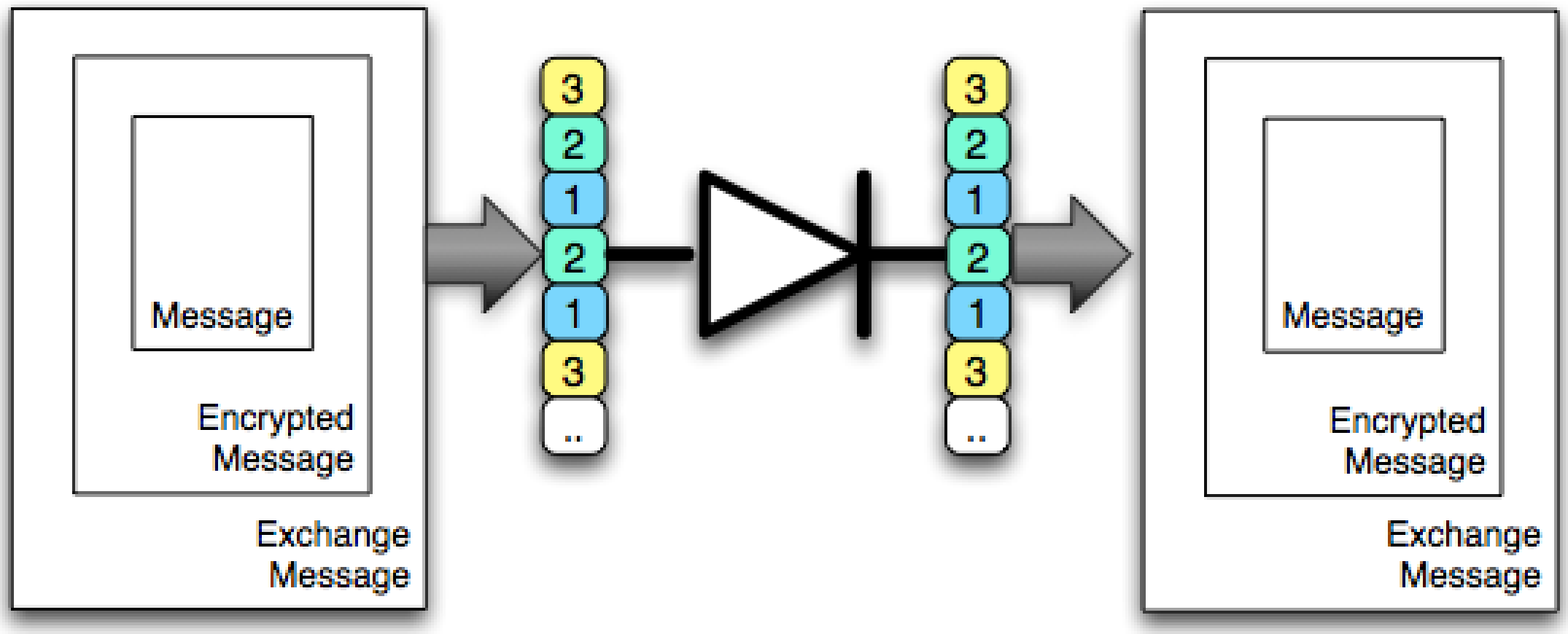

Fig 18: Encrypted Message in Exchange Message

## 5. Description of results

With the application it is now possible to send messages of arbitrary size over a diode where exchanges are automatically mirrored and compression is used to maximise the data in a message. Below follows a more detailed explanation of the results.

### Test setup

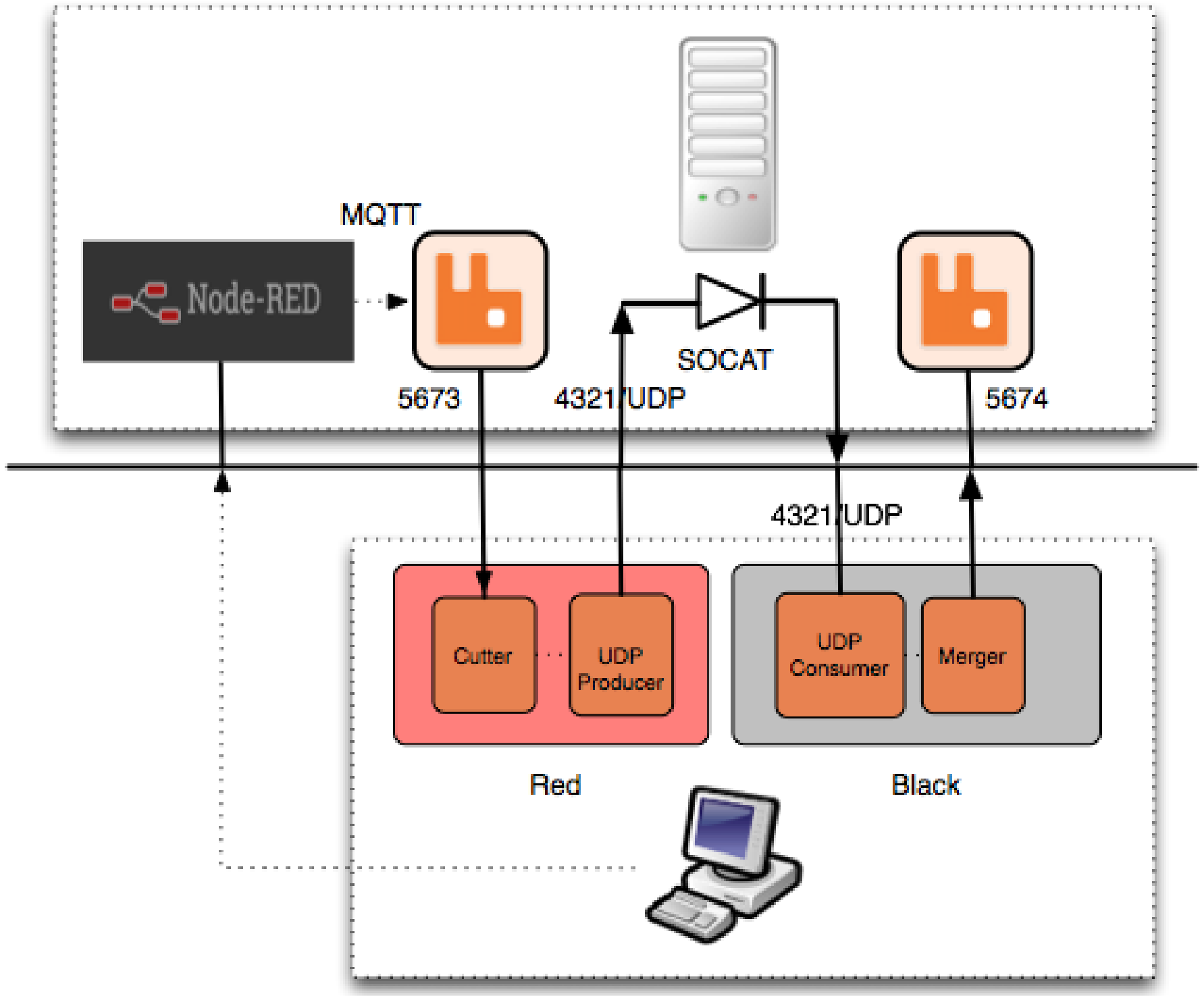

Fig 19: test setup

The network is a gigabit network connecting a workstation running the black and red processes and a server running both RabbitMQs and Node-RED and a virtual data diode with socat:

```
$ socat UDP4-RECVFROM:4321,fork UDP4-SENDTO:<<IP_OF_WORKSTATION>>:4321
```

**Semi-production setup**

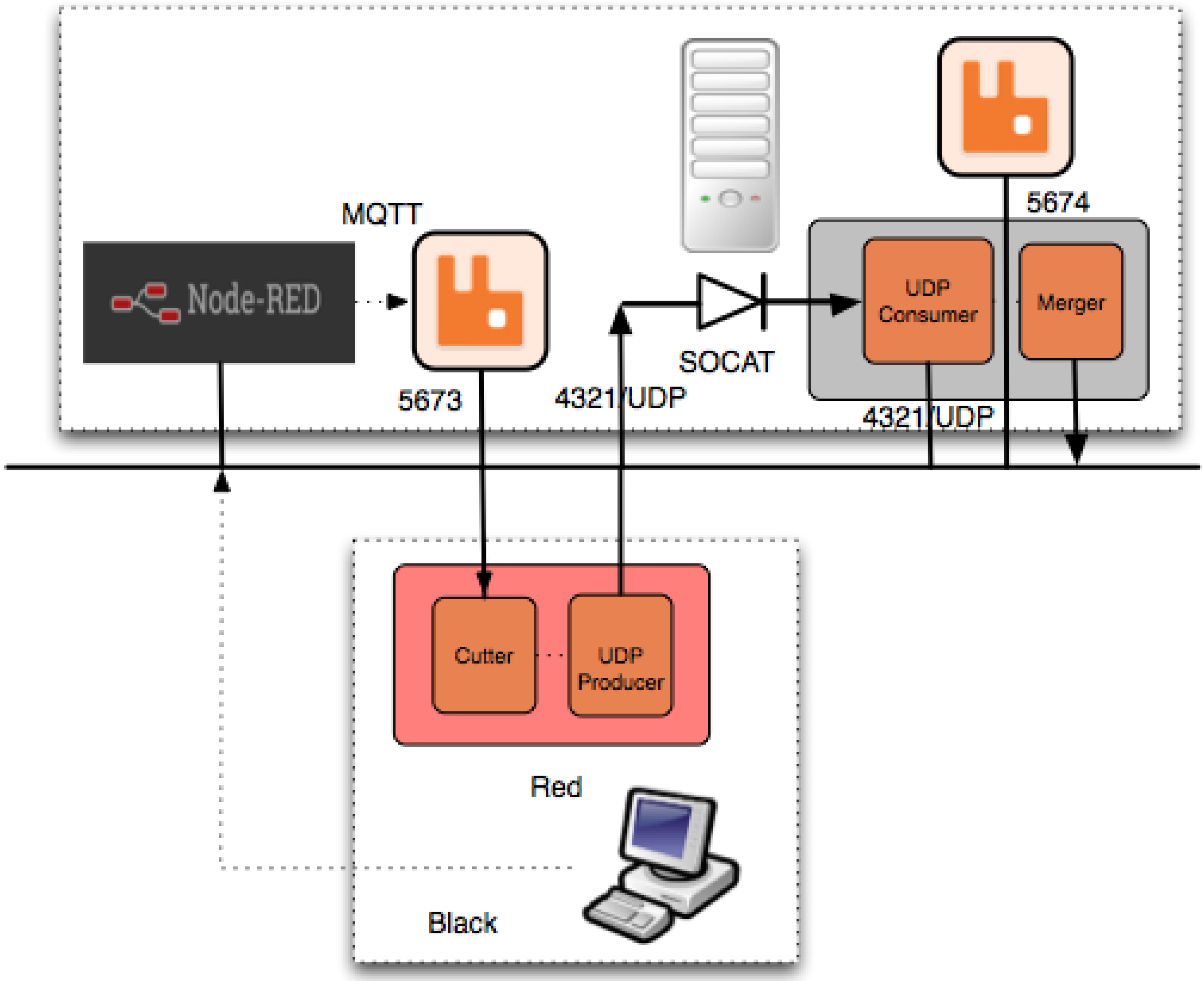

Fig 20: Semi-production setup

**Production setup**

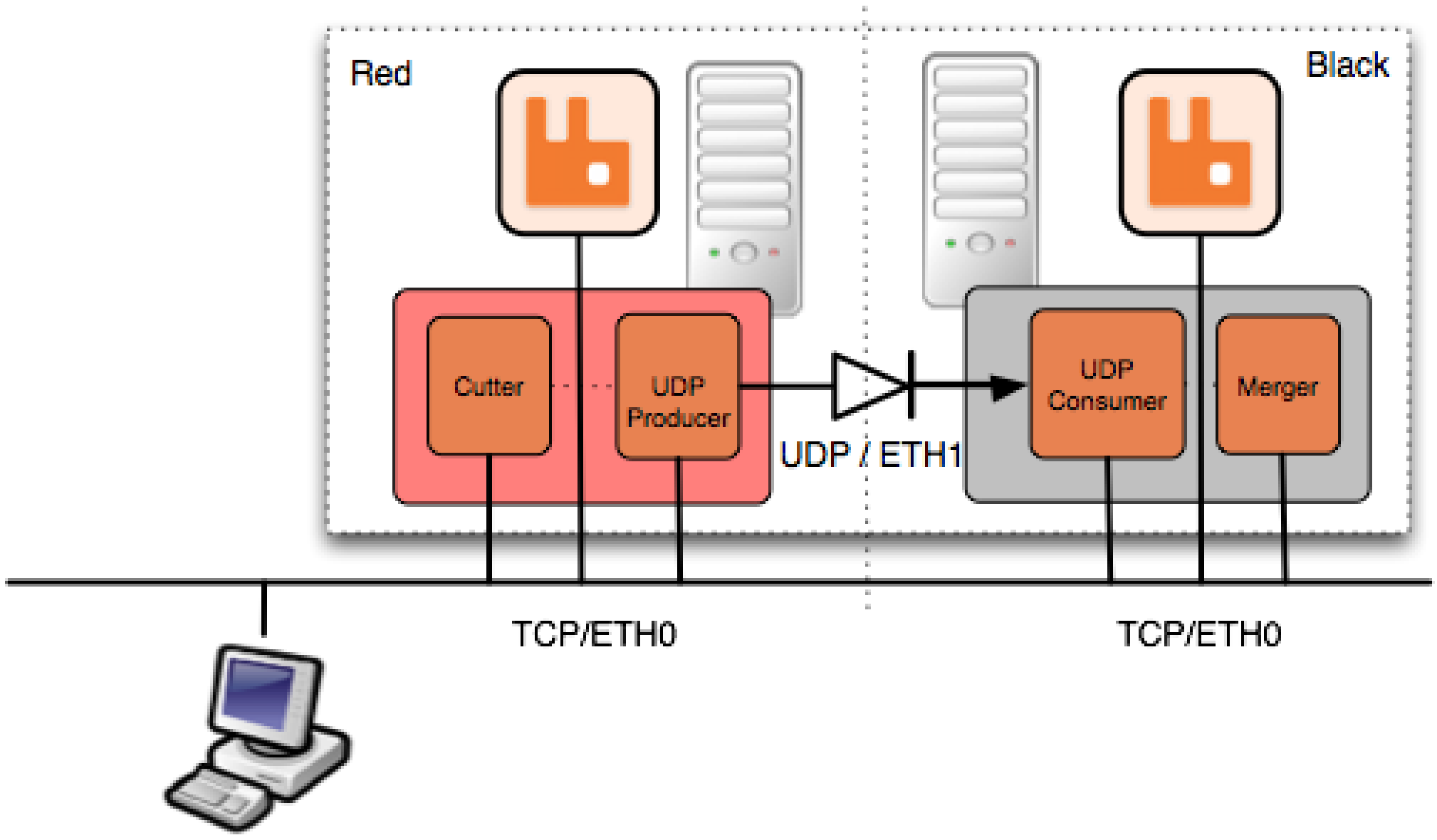

Fig 21: Production setup

**Automatic creation of exchanges**

The following are screenshots that show the application where a fictitious sensor which is declared in the class "TemperatureSensor" sends periodically TemperatureSensorEvent events over to the red side. Without any management exchanges and queues are created automatically and the data lands over a data diode into the trusted zone:

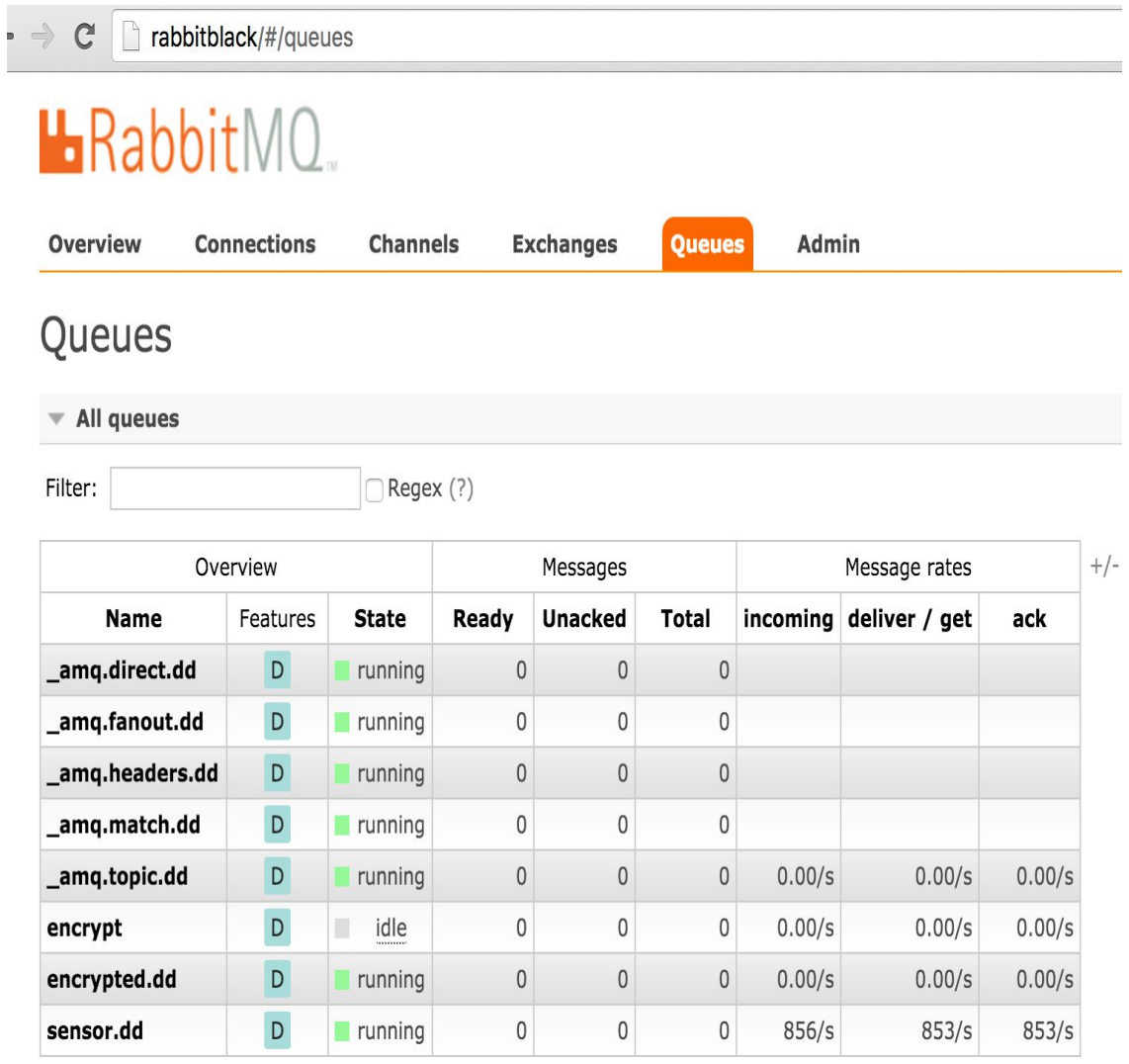

Black exchanges

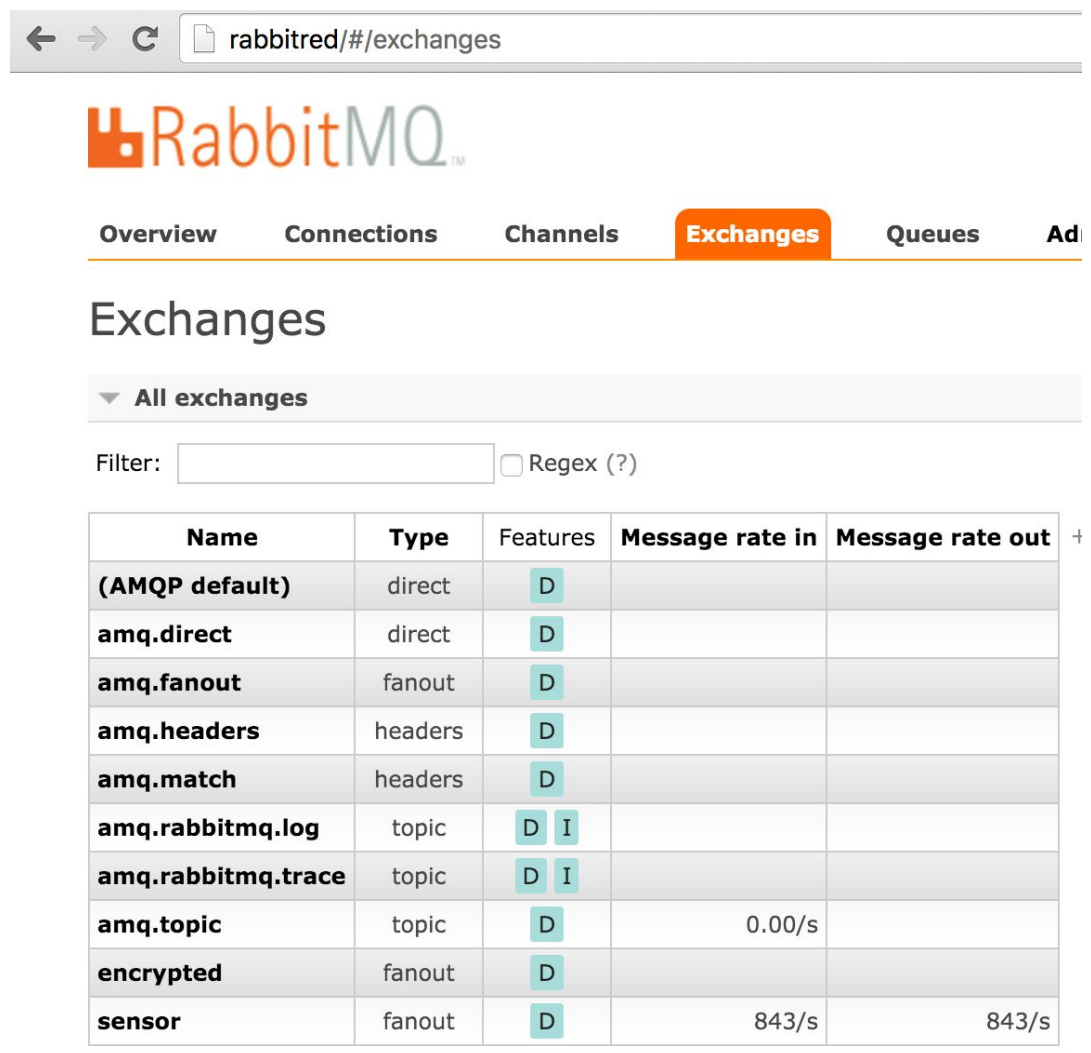

Red exchanges

Message rate has been measured of over 14.500 messages/sec on a laptop running both black and red applications and and gigabit ethernet between the laptop and a Intel NUC running both RabbitMQs and the Node-RED applications.

**Performance**

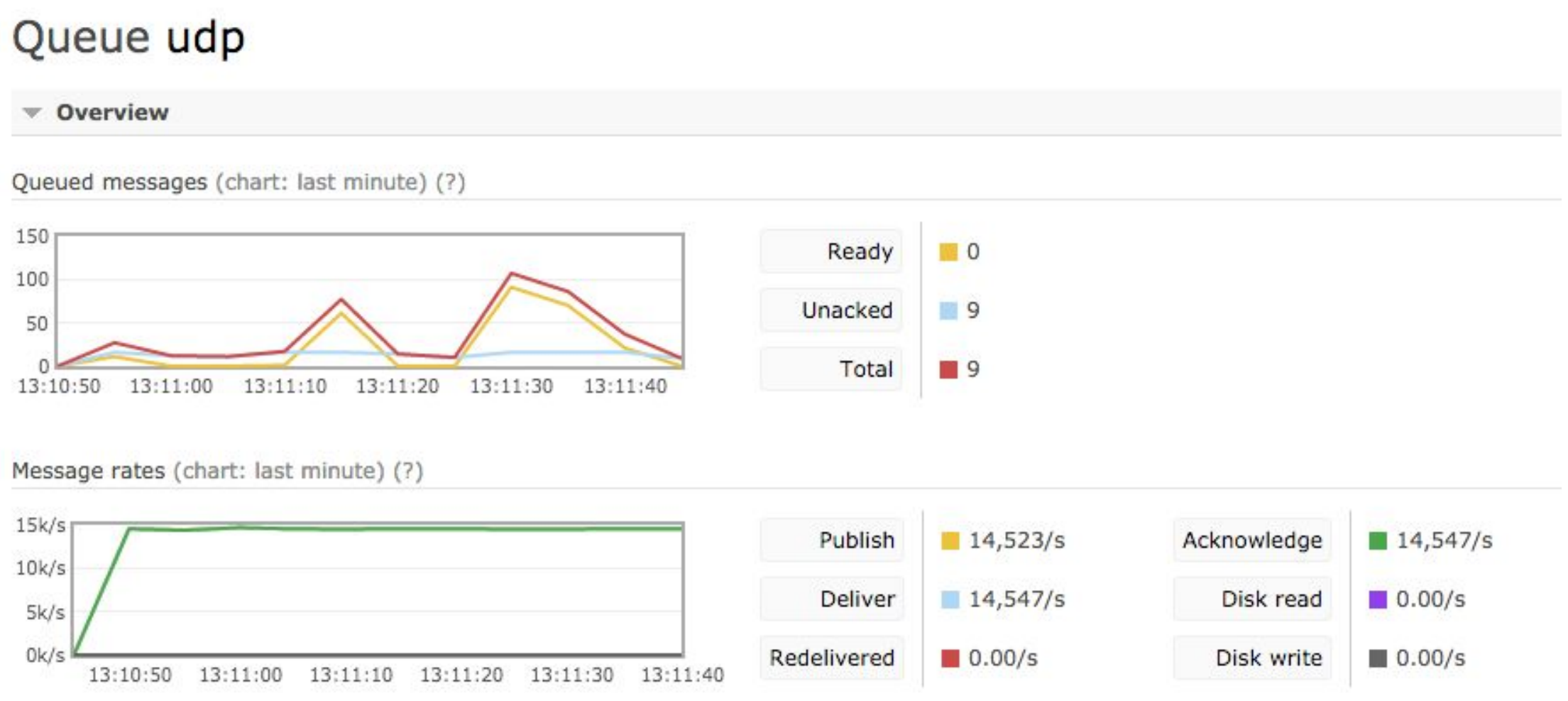

Fig 22: graphical throughput of UDP receiver

In the above figure the actual throughput over UDP can be observed.

On the black side a script “gradle runClient” is started which sends packets with random contents over UDP to the red side. In each packet the first 4 bytes contain an index number and each packet increments the index. On the receiving side a dedicated C program (“udp”) takes a UDP message and puts into the exchange UDP. Then a second process (“gradle runServer”) receives the message and the index is compared to the previous received index. If the index does not equal the previous index + 1 the server outputs an error message. The rate at which no packets are lost can thus be determined. The highest message rate where no packets are lost is a rate of 14.500 packets/sec with packet size of 8K which is (14.500 * 8192)/1024/1024 = 113MB/sec which is 89% of the theoretical limit of of Gigabit ethernet at 128M/sec.

Now the practical test: the script datadiode/contrib/nodejs/sendMessages.js generates 1MB messages in the exchange “nodejsExchange” which are then automatically transferred to the red side. Because those messages are exactly 1MB in size the message rate reflects the throughput in MB/sec.

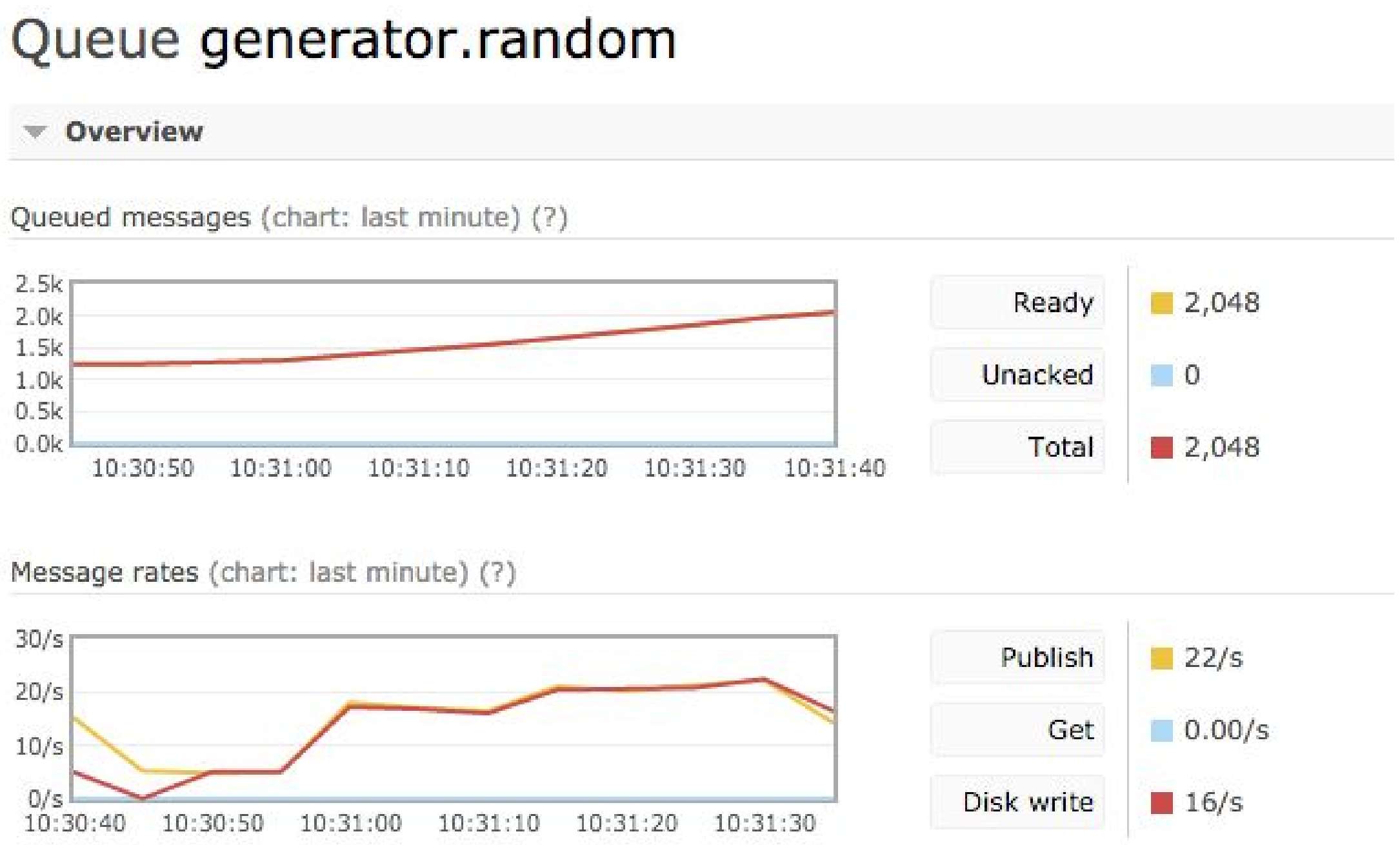

Fig 23: Measured throughput of 1MB messages

The best performance was found with 8K jumbo frames enabled currently giving a throughput of around 25MB/sec with no packet loss. Because packets are sent twice (redundancy factor is 2 in this example) the actual throughput is around 50MB/sec which is far below the 128MB/sec maximum of a gigabit network. The reason it does not perform as fast as the previous test is the cut and merge function both sides have to do which cannot put packets faster than 700 packets/sec and message digest calculations both sides have to do. This can be solved by performance tuning both the cut and merge applications and to put those applications on different computer(s) connected to both RabbitMQs instead of running all these processes on a single computational instance as is done in this example.

## MQTT

These screenshots show the Node-RED applications where MQTT messages are successfully being transferred over the data diode.

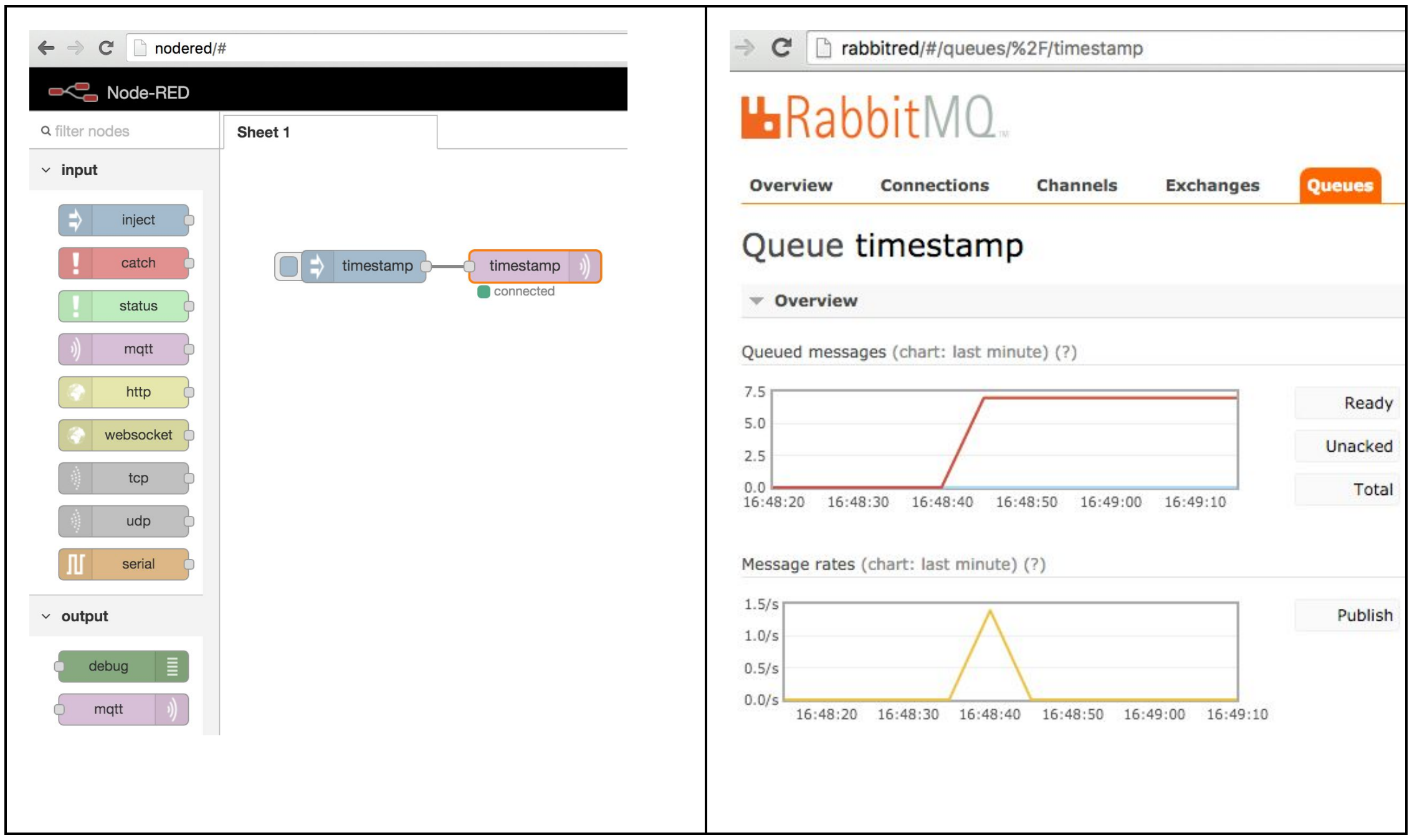

Fig 24: Example of MQTT over the data diode via Node-RED

Other that declaring a binding for the routing key because MQTT sends to the Topic exchange, no further modifications had to made to support MQTT. More screenshots can be found in the appendix on how to start the applications to reproduce the results.

**Compression**

Wireshark analysis showed that the serialized TemperatureSensorEvents, used to mimic a real life sensor, are 2378 bytes:

```
▽ User Datagram Protocol, Src Port: 35138 (35138), Dst Port: 1234 (1234)
    Source Port: 35138 (35138)
    Destination Port: 1234 (1234)
    Length: 2386
  ▷ Checksum: 0xd27d [validation disabled]
    [Stream index: 8]
▽ Data (2378 bytes)
    Data: aced00057372002b6f72672e6461746164696f64652e6d6f...
    [Length: 2378]
```

Visual inspection of the data showed that there is quite some text and compression and decompression in the form of Inflater and Deflater could be used to reduce the packet size. An field in application.properties called

```
application.datadiode.black.udp.compress = true
```

can switch off or on compression and showed the following result:

| | body in bytes | Message in bytes | Exchange Message | compressed in bytes | reduction |
|---|---|---|---|---|---|
| temperatuur sensor | 810 | 2041 | 2378 | 1247 | 52 % |
| MQTT message | 13 | 1297 | 1636 | 927 | 56 % |

With compression, messages show a 50% reduction.

## Encryption

After that we apply encryption to the sensor-stream. As we can see packets arrive at around 750 messages/sec at the black side which are then encrypted and sent over the data diode. This happens at a rate of 100 messages/sec. On the red side those messages are validated and decrypted and arrive at the speed of around 100 messages/sec in the sensor-exchange on the red side.

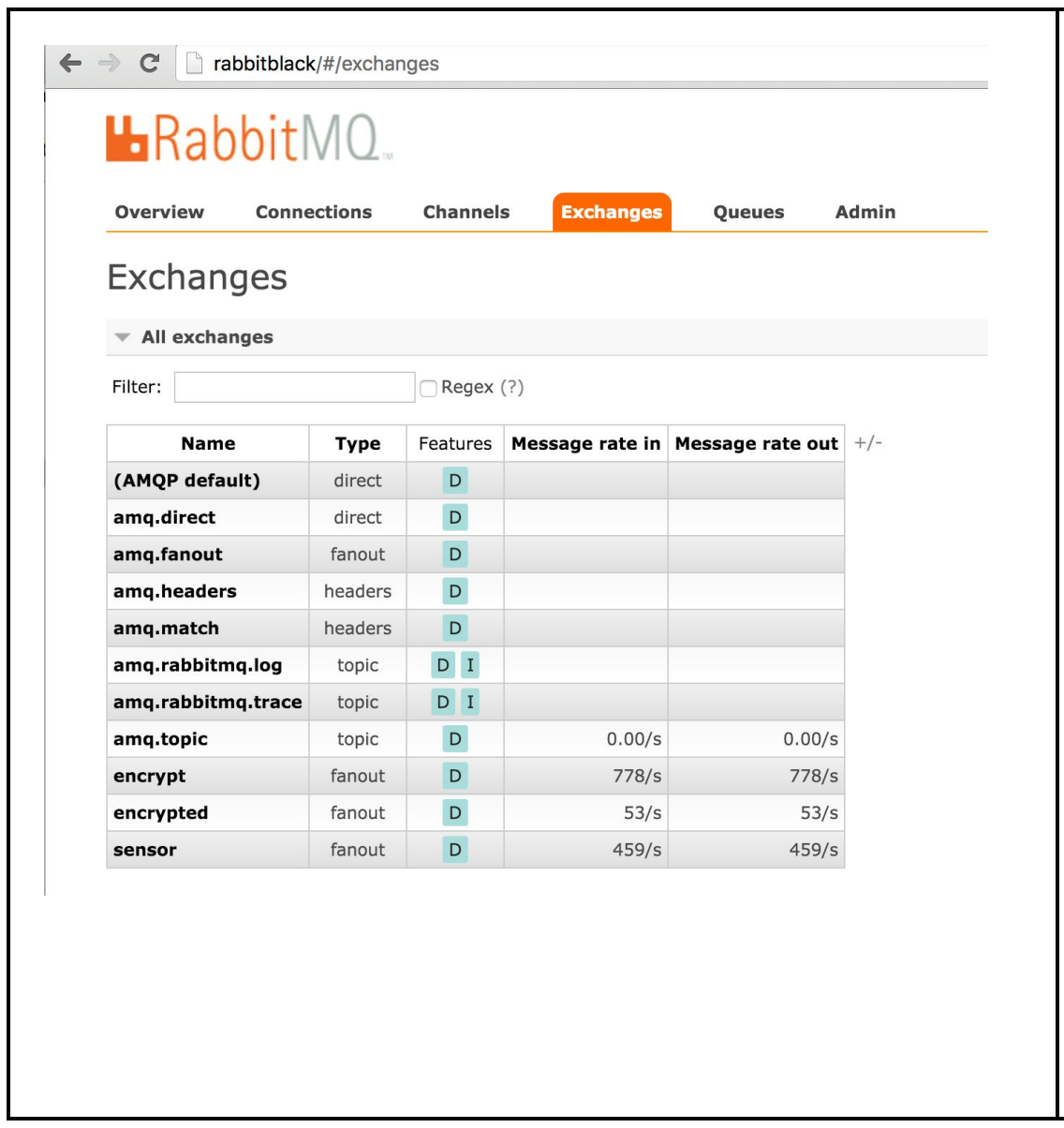

Black exchanges

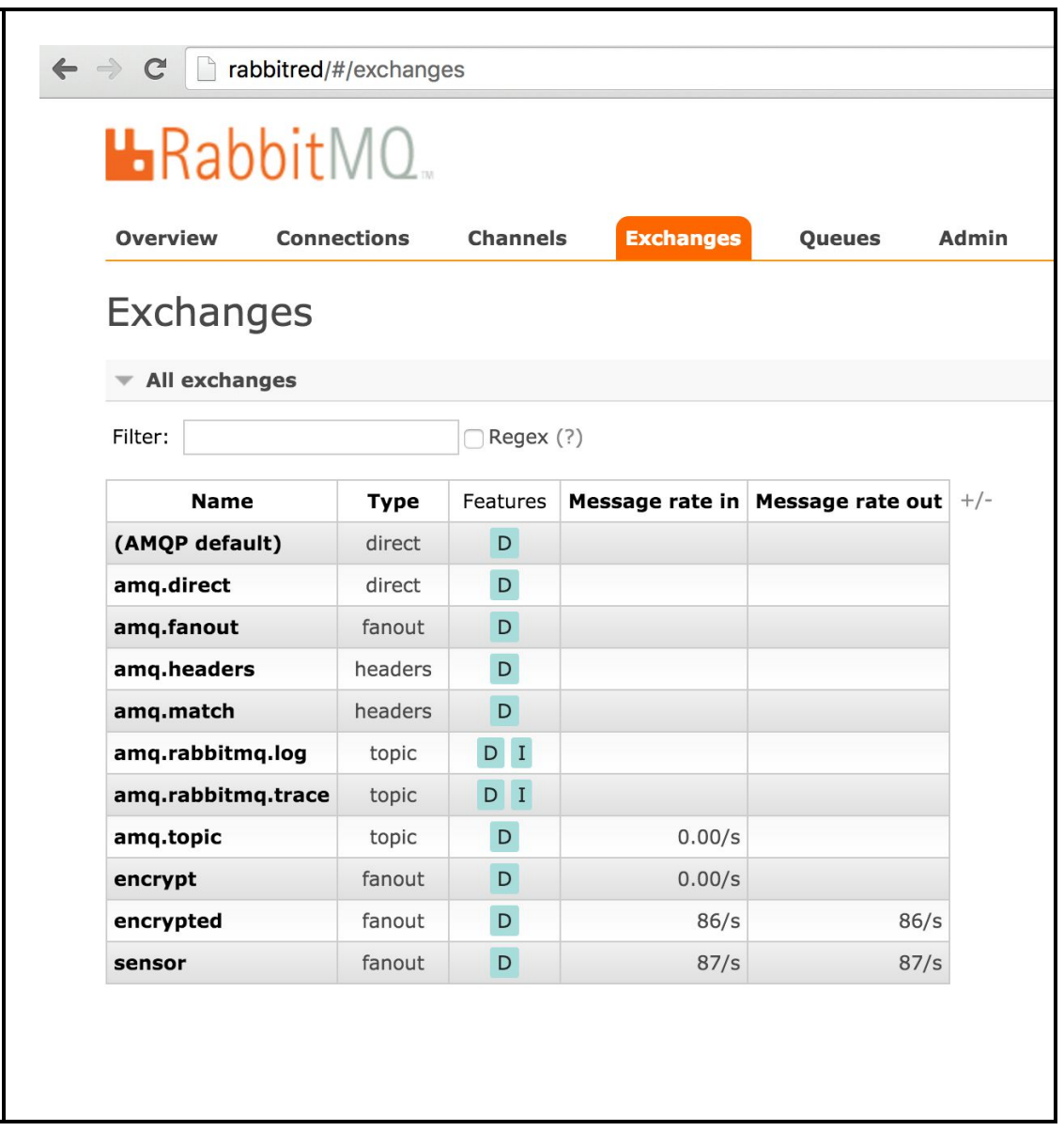

Red exchanges

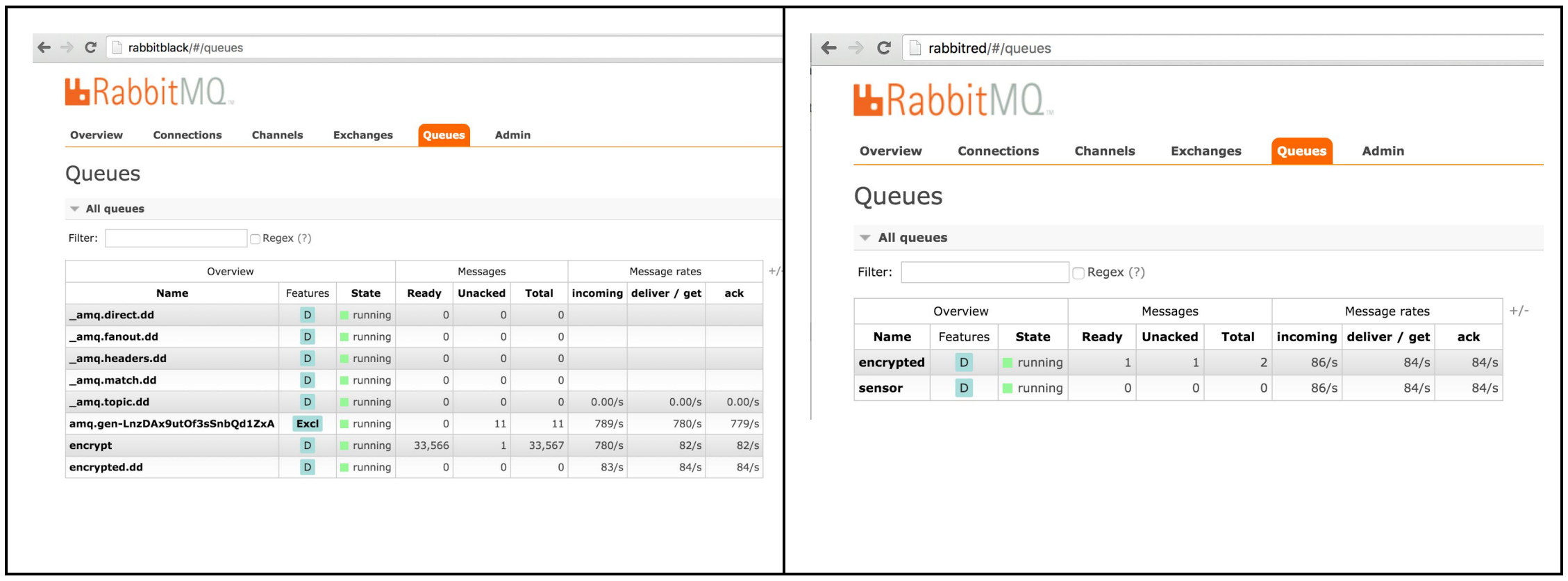

Black queues Red queues

The encryption process cannot keep up because messages arrive at a faster rate than the encryption process can encrypt them. Normally messages would be lost when a process cannot keep up but messages are now stored in the encrypt queue until they can be processed. A solution would be to start more encryption processes and because as shown compression happens at 100 messages/sec and messages arrive at around 750 messages/sec we would probably need 7 more encryption processes.

**Missing messages**

On the red side missing messages are detected at the SensorListener. The TemperatureSensor sends periodically TemperatureSensorEvent to the red side. Since it inherits from Event is has a field "index" which is an incrementing counter. At the red side it is now trivial to measure missing packets: if the last index value is not current index counter value added by one, a warning is being displayed in the logfile.

**Discussion of results**

The developed application for this dissertation provides a working reference implementation of a mirrored RabbitMQ over a data diode and includes all the scripts used to measure the performance. The results of this application show that it possible to transfer data of arbitrary message length over a data diode and further shows that encryption can successfully be used for messages containing sensitive information. Exchanges are automatically created and unencrypted and encrypted messages can successfully be mirrored and decrypted over a data diode. Everything is maintenance-free, except for key-management if encryption is used. The transferred messages can be of arbitrary size because bigger messages are split into smaller segments before they are transferred over the diode and each transferred message is validated by a checksum.

Evaluation of the results thus show that it is definitely worth to do further research to increase the throughput even further than the currently measured 25MB/sec but in its current form the application is successfully deployed into production.

Follow-on work in this area is to performance tune the cut and merge applications and to use a database as a backend for the reconstruction of messages. The current setup is more of a proof of concept: it works but does all the reconstruction in memory. This has the disadvantage that a restart of the application means loss of data and also means that the size of messages cannot be bigger than the amount of free memory and will enable the possibility of reconstructing messages in parallel and even implement priority queues.

Also the current implementation sends packets data twice over the diode to minimise the impact of packet loss. This could also be accomplished by putting Forward Error Correction blocks in the stream, where the result of XOR ing blocks can be used to reconstruct a missing block but has the advantage that the redundancy factor can be minimised by sending 4 blocks data and 1 block FEC instead of sending each block twice.

Follow-on work also includes the use of multiple decryption keys for a single message with the help of gnupg so that a single message could be decrypted by multiple recipients.

## 6. List of references

## 7. Appendices

All of the source code for the reference application can be found on github:

- RabbitMQ Applications: “Data Diode”

Check the code out with git on both nodes and start the infrastructure with docker-compose:

```
marcel@docker-01:~/projects/docker-spring-datadiode-black/contrib/docker$ docker-compose up
Creating docker_proxy_1...
Creating docker_volumes_1...
Creating docker_rabbitred_1...
Creating docker_rabbitblack_1...
Creating docker_nodered_1...
Attaching to docker_proxy_1, docker_rabbitred_1, docker_rabbitblack_1, docker_nodered_1
```

Tune the network parameters for high performance throughput

```
#!/bin/bash
ifconfig p1p1 192.168.1.1 mtu 9710 txqueuelen 1000 up
arp -s 192.168.1.2 <<INTERFACE MAC ADDRESS RED>>
ip route add default via 192.168.1.1
sysctl -w \
 net.core.wmem_max=26214400 \
 net.core.wmem_default=8388608 \
 fs.file-max=100000 \
 vm.swappiness=10 \
 net.core.optmem_max=40960 \
 net.core.netdev_max_backlog=50000 \
 net.ipv4.udp_mem='8388608 8388608 8388608' \
 net.ipv4.udp_wmem_min=8388608 \
 net.ipv4.conf.all.send_redirects=0 \
 net.ipv4.conf.all.accept_redirects=0 \
 net.ipv4.conf.all.accept_source_route=0 \
 net.ipv4.conf.all.log_martians=1 \
 net.netfilter.nf_conntrack_max=262144 \
 net.nf_conntrack_max=262144
```

network_tune.sh: tune black side (sender)

```
#!/bin/bash
ifconfig p1p1 192.168.1.2 mtu 9710
arp -s 192.168.1.1 <<INTERFACE MAC ADDRESS BLACK>>
sudo sysctl -w \
 net.core.rmem_max=86214400 \
 net.core.rmem_default=46214400 \
 fs.file-max=2097152 \
 vm.swappiness=10 \
 net.core.optmem_max=80960 \
 net.core.netdev_max_backlog=70000 \
 net.ipv4.udp_mem='26214400 26214400 26214400' \
 net.ipv4.udp_rmem_min=26214400 \
 net.ipv4.conf.all.send_redirects=0 \
 net.ipv4.conf.all.accept_redirects=0 \
 net.ipv4.conf.all.accept_source_route=0 \
 net.ipv4.conf.all.log_martians=1 \
 net.netfilter.nf_conntrack_max=262144 \
 net.nf_conntrack_max=262144
```

network_tune.sh: tune red side (receiver)

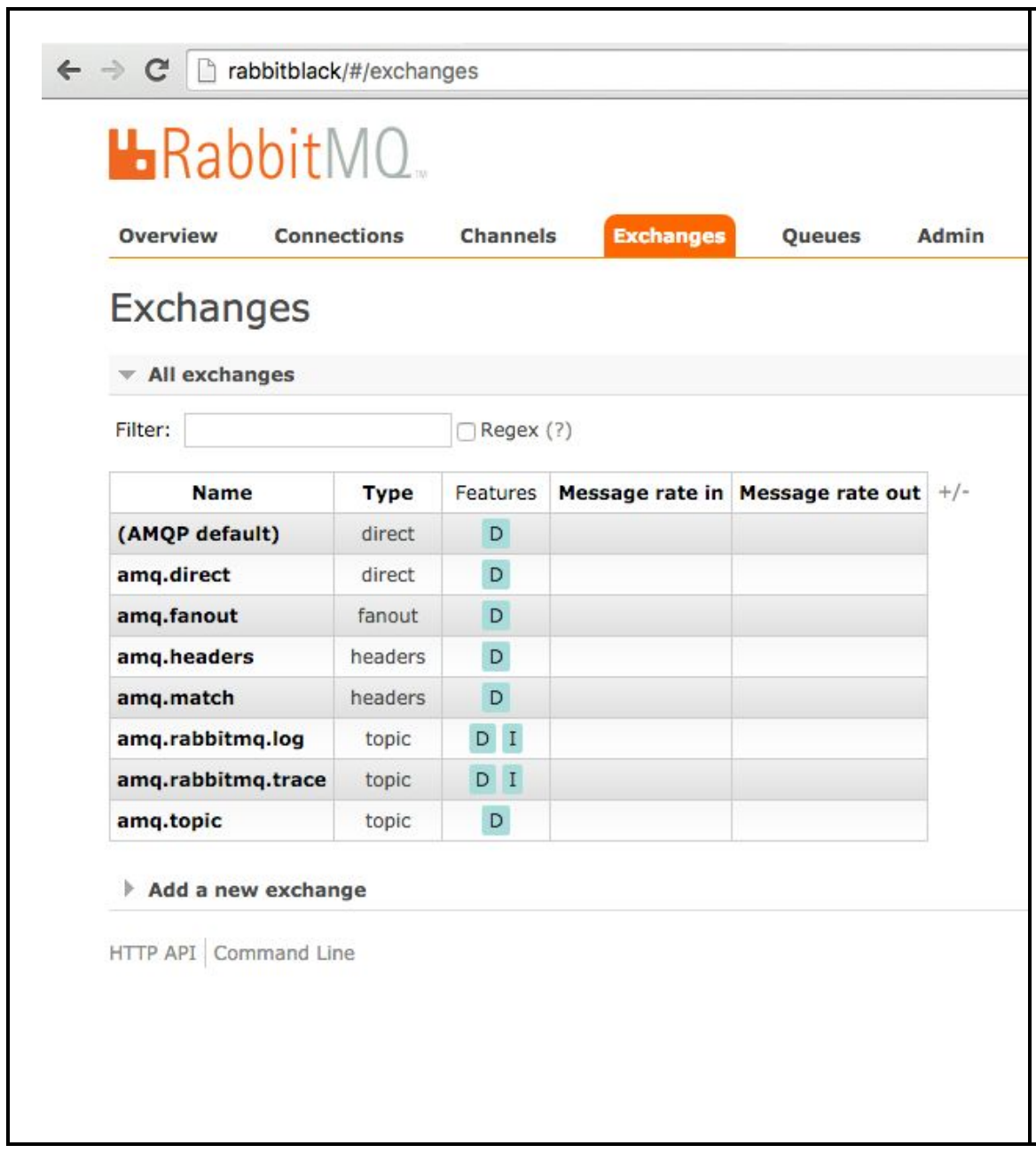

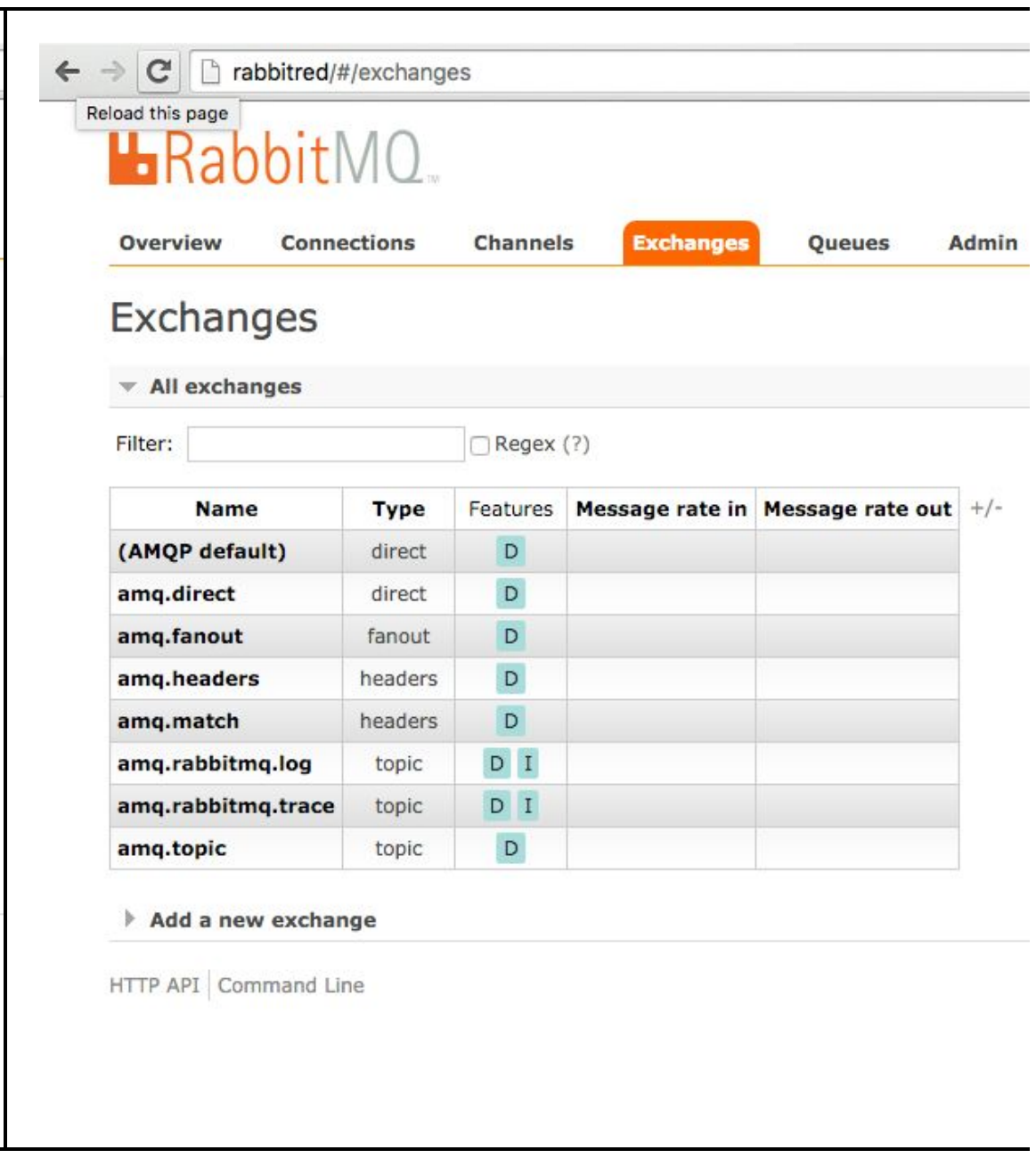

[http://rabbitblack/#/exchanges](http://rabbitblack/#/exchanges) [http://rabbitred/#/exchanges](http://rabbitred/#/exchanges)

START BLACK

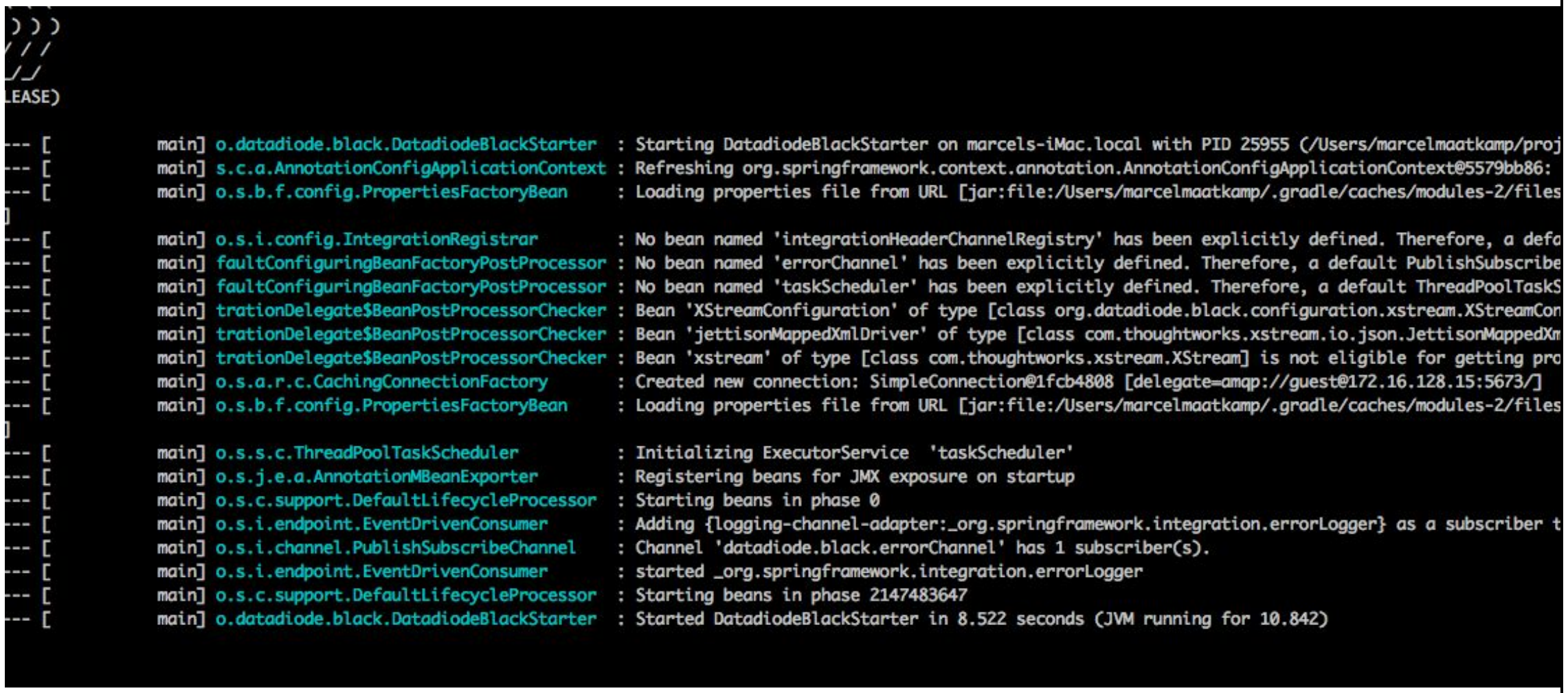

```
) ) )
/ / /
_/_/
LEASE)

--- [          main] o.datadiode.black.DatadiodeBlackStarter  : Starting DatadiodeBlackStarter on marcels-iMac.local with PID 25955 (/Users/marcelmaatkamp/proj
--- [          main] s.c.a.AnnotationConfigApplicationContext : Refreshing org.springframework.context.annotation.AnnotationConfigApplicationContext@5579bb86:
--- [          main] o.s.b.f.config.PropertiesFactoryBean     : Loading properties file from URL [jar:file:/Users/marcelmaatkamp/.gradle/caches/modules-2/files
]
--- [          main] o.s.i.config.IntegrationRegistrar        : No bean named 'integrationHeaderChannelRegistry' has been explicitly defined. Therefore, a defa
--- [          main] faultConfiguringBeanFactoryPostProcessor : No bean named 'errorChannel' has been explicitly defined. Therefore, a default PublishSubscribe
--- [          main] faultConfiguringBeanFactoryPostProcessor : No bean named 'taskScheduler' has been explicitly defined. Therefore, a default ThreadPoolTaskS
--- [          main] trationDelegate$BeanPostProcessorChecker : Bean 'XStreamConfiguration' of type [class org.datadiode.black.configuration.xstream.XStreamCon
--- [          main] trationDelegate$BeanPostProcessorChecker : Bean 'jettisonMappedXmlDriver' of type [class com.thoughtworks.xstream.io.json.JettisonMappedXm
--- [          main] trationDelegate$BeanPostProcessorChecker : Bean 'xstream' of type [class com.thoughtworks.xstream.XStream] is not eligible for getting pro
--- [          main] o.s.a.r.c.CachingConnectionFactory       : Created new connection: SimpleConnection@1fcb4808 [delegate=amqp://guest@172.16.128.15:5673/]
--- [          main] o.s.b.f.config.PropertiesFactoryBean     : Loading properties file from URL [jar:file:/Users/marcelmaatkamp/.gradle/caches/modules-2/files
]
--- [          main] o.s.s.c.ThreadPoolTaskScheduler          : Initializing ExecutorService  'taskScheduler'
--- [          main] o.s.j.e.a.AnnotationMBeanExporter        : Registering beans for JMX exposure on startup
--- [          main] o.s.c.support.DefaultLifecycleProcessor  : Starting beans in phase 0
--- [          main] o.s.i.endpoint.EventDrivenConsumer       : Adding {logging-channel-adapter:_org.springframework.integration.errorLogger} as a subscriber t
--- [          main] o.s.i.channel.PublishSubscribeChannel    : Channel 'datadiode.black.errorChannel' has 1 subscriber(s).
--- [          main] o.s.i.endpoint.EventDrivenConsumer       : started _org.springframework.integration.errorLogger
--- [          main] o.s.c.support.DefaultLifecycleProcessor  : Starting beans in phase 2147483647
--- [          main] o.datadiode.black.DatadiodeBlackStarter  : Started DatadiodeBlackStarter in 8.522 seconds (JVM running for 10.842)
```

START RED

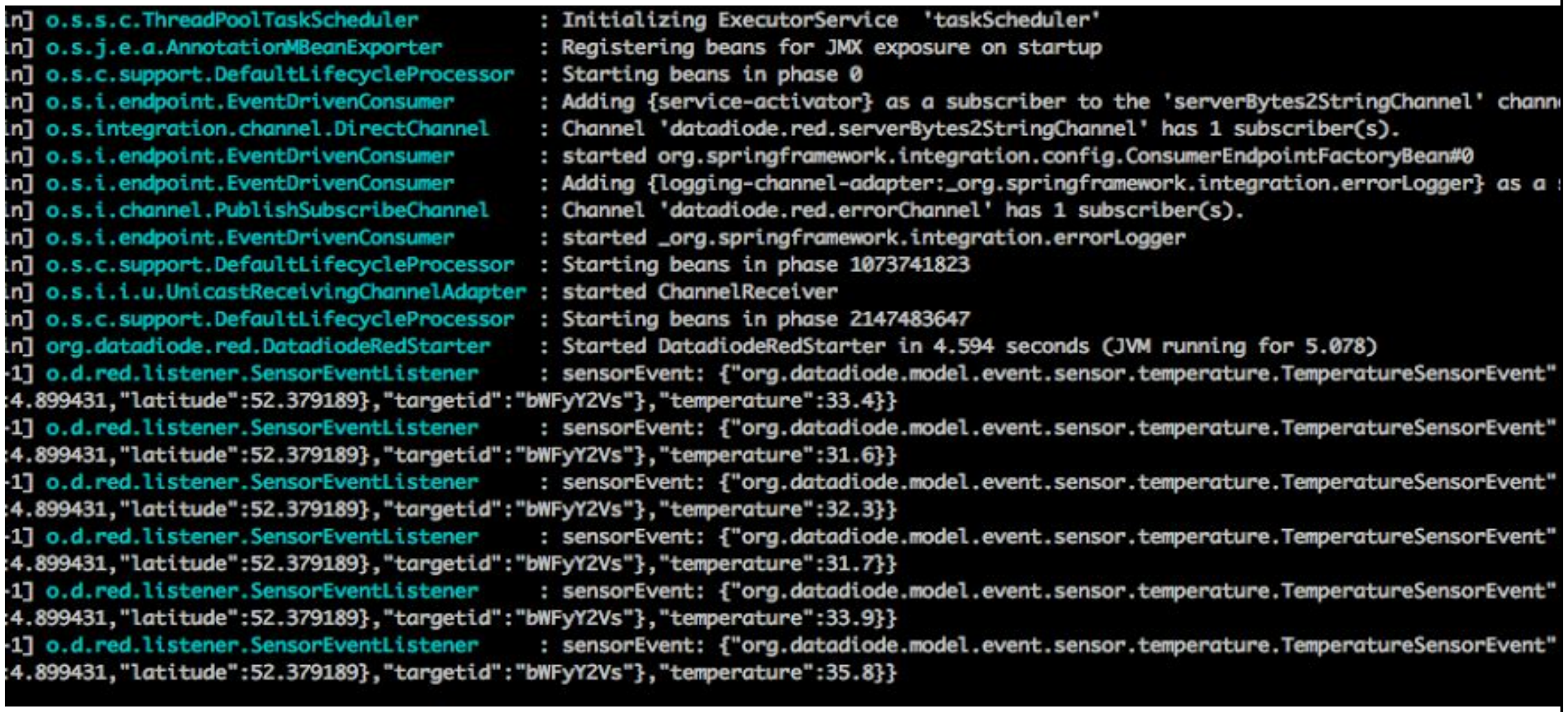

```
in] o.s.s.c.ThreadPoolTaskScheduler          : Initializing ExecutorService  'taskScheduler'
in] o.s.j.e.a.AnnotationMBeanExporter        : Registering beans for JMX exposure on startup
in] o.s.c.support.DefaultLifecycleProcessor  : Starting beans in phase 0
in] o.s.i.endpoint.EventDrivenConsumer       : Adding {service-activator} as a subscriber to the 'serverBytes2StringChannel' chann
in] o.s.integration.channel.DirectChannel    : Channel 'datadiode.red.serverBytes2StringChannel' has 1 subscriber(s).
in] o.s.i.endpoint.EventDrivenConsumer       : started org.springframework.integration.config.ConsumerEndpointFactoryBean#0
in] o.s.i.endpoint.EventDrivenConsumer       : Adding {logging-channel-adapter:_org.springframework.integration.errorLogger} as a
in] o.s.i.channel.PublishSubscribeChannel    : Channel 'datadiode.red.errorChannel' has 1 subscriber(s).
in] o.s.i.endpoint.EventDrivenConsumer       : started _org.springframework.integration.errorLogger
in] o.s.c.support.DefaultLifecycleProcessor  : Starting beans in phase 1073741823
in] o.s.i.i.u.UnicastReceivingChannelAdapter : started ChannelReceiver
in] o.s.c.support.DefaultLifecycleProcessor  : Starting beans in phase 2147483647
in] org.datadiode.red.DatadiodeRedStarter    : Started DatadiodeRedStarter in 4.594 seconds (JVM running for 5.078)
-1] o.d.red.listener.SensorEventListener     : sensorEvent: {"org.datadiode.model.event.sensor.temperature.TemperatureSensorEvent"
:4.899431,"latitude":52.379189},"targetid":"bWFyY2Vs"},"temperature":33.4}}
-1] o.d.red.listener.SensorEventListener     : sensorEvent: {"org.datadiode.model.event.sensor.temperature.TemperatureSensorEvent"
:4.899431,"latitude":52.379189},"targetid":"bWFyY2Vs"},"temperature":31.6}}
-1] o.d.red.listener.SensorEventListener     : sensorEvent: {"org.datadiode.model.event.sensor.temperature.TemperatureSensorEvent"
:4.899431,"latitude":52.379189},"targetid":"bWFyY2Vs"},"temperature":32.3}}
-1] o.d.red.listener.SensorEventListener     : sensorEvent: {"org.datadiode.model.event.sensor.temperature.TemperatureSensorEvent"
:4.899431,"latitude":52.379189},"targetid":"bWFyY2Vs"},"temperature":31.7}}
-1] o.d.red.listener.SensorEventListener     : sensorEvent: {"org.datadiode.model.event.sensor.temperature.TemperatureSensorEvent"
:4.899431,"latitude":52.379189},"targetid":"bWFyY2Vs"},"temperature":33.9}}
-1] o.d.red.listener.SensorEventListener     : sensorEvent: {"org.datadiode.model.event.sensor.temperature.TemperatureSensorEvent"
:4.899431,"latitude":52.379189},"targetid":"bWFyY2Vs"},"temperature":35.8}}
```

Verify that a sensor sends its messages periodically to the black side, which are mirrored over to the red side.

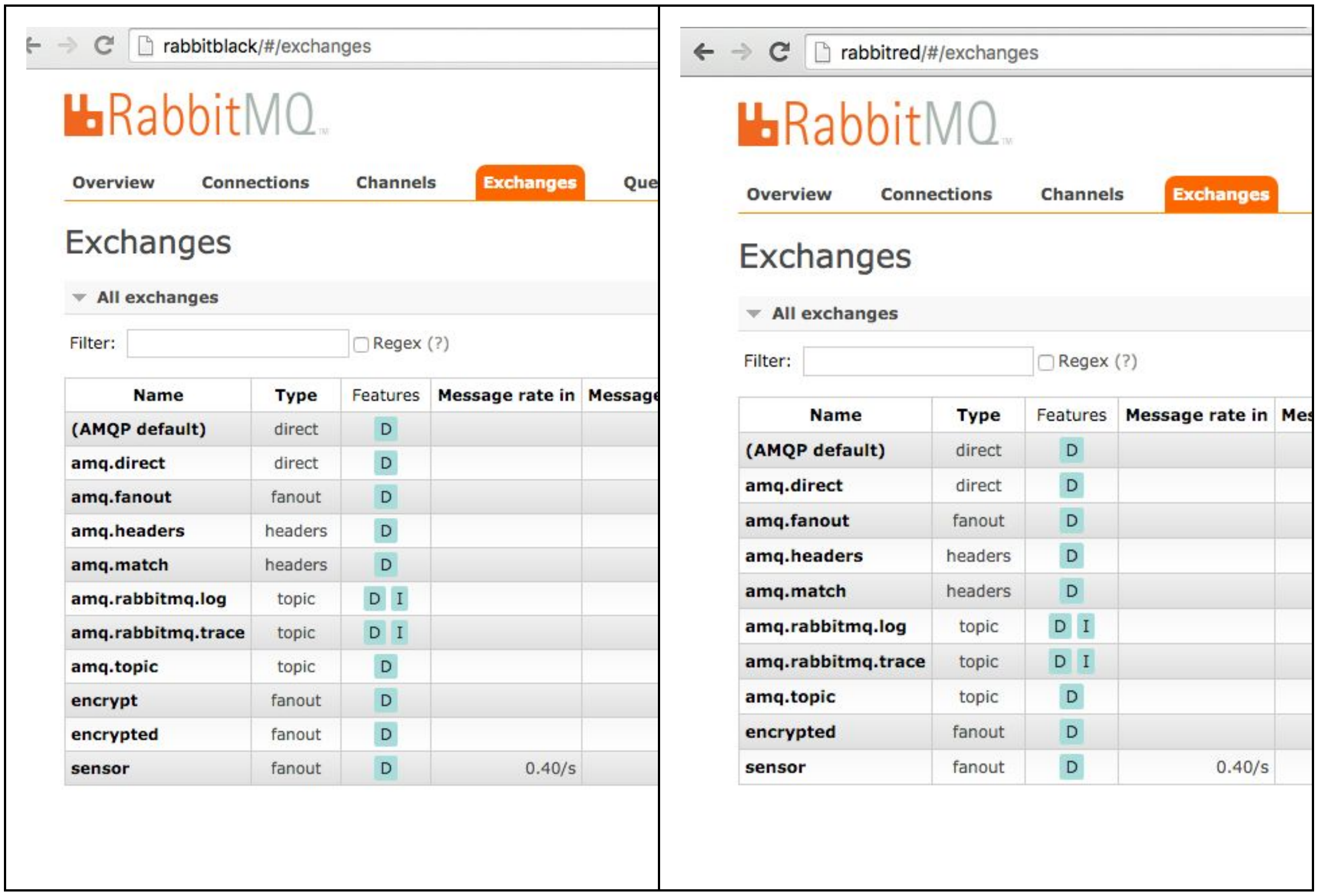

[http://rabbitblack/#/exchanges](http://rabbitblack/#/exchanges) [http://rabbitred/#/exchanges](http://rabbitred/#/exchanges)

Go to the Node-RED instance and add a ‘Inject’ input-node and call it “timestamp” which will send a timestamp every time the button is pushed and add it to a MQTT output node where the broker is the black side RabbitMQ.

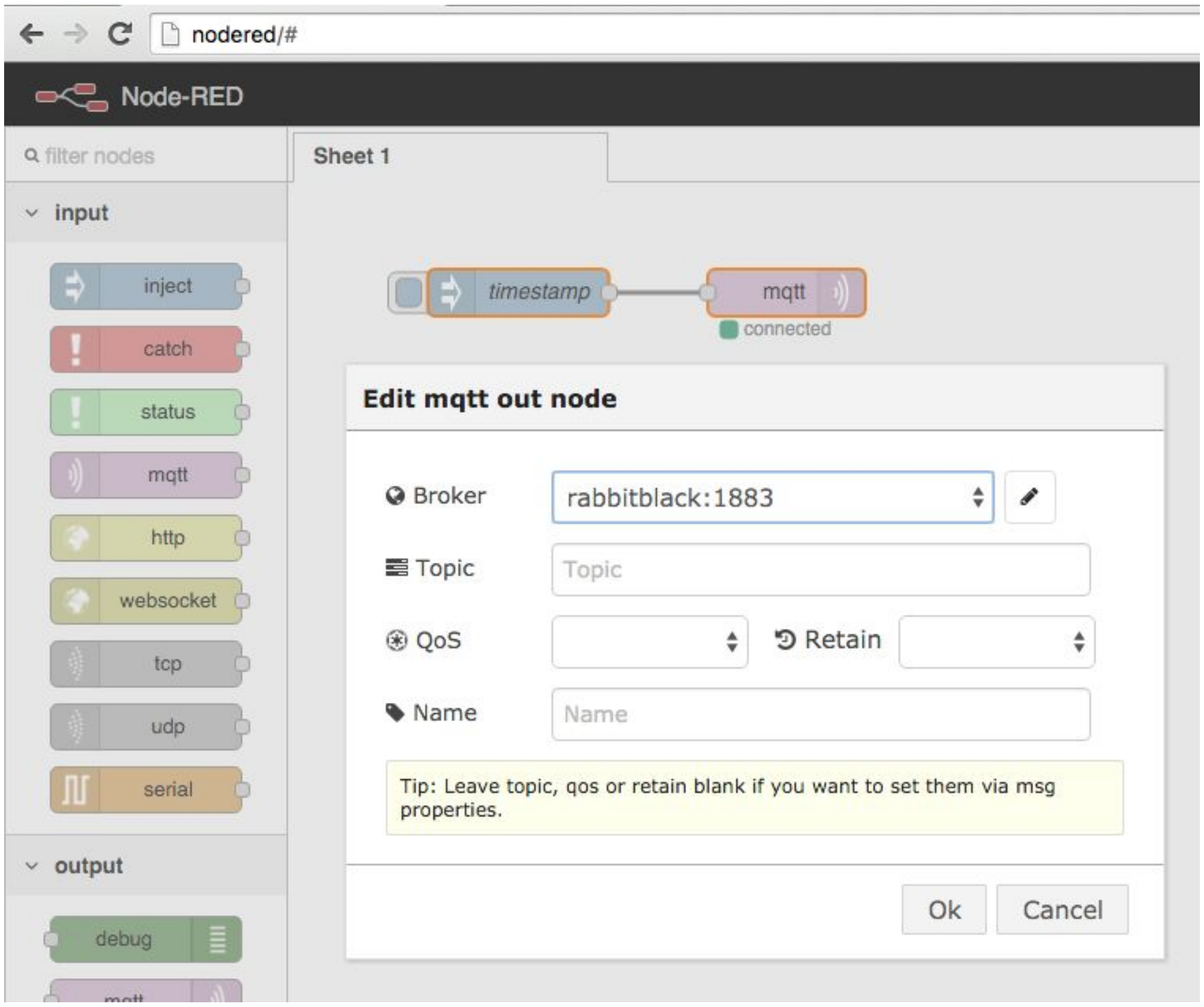

[http://nodered/#](http://nodered/#)

Test the setup by pushing a few times on the button in Node-RED and see that messages are indeed being send from the black side over to the red side.

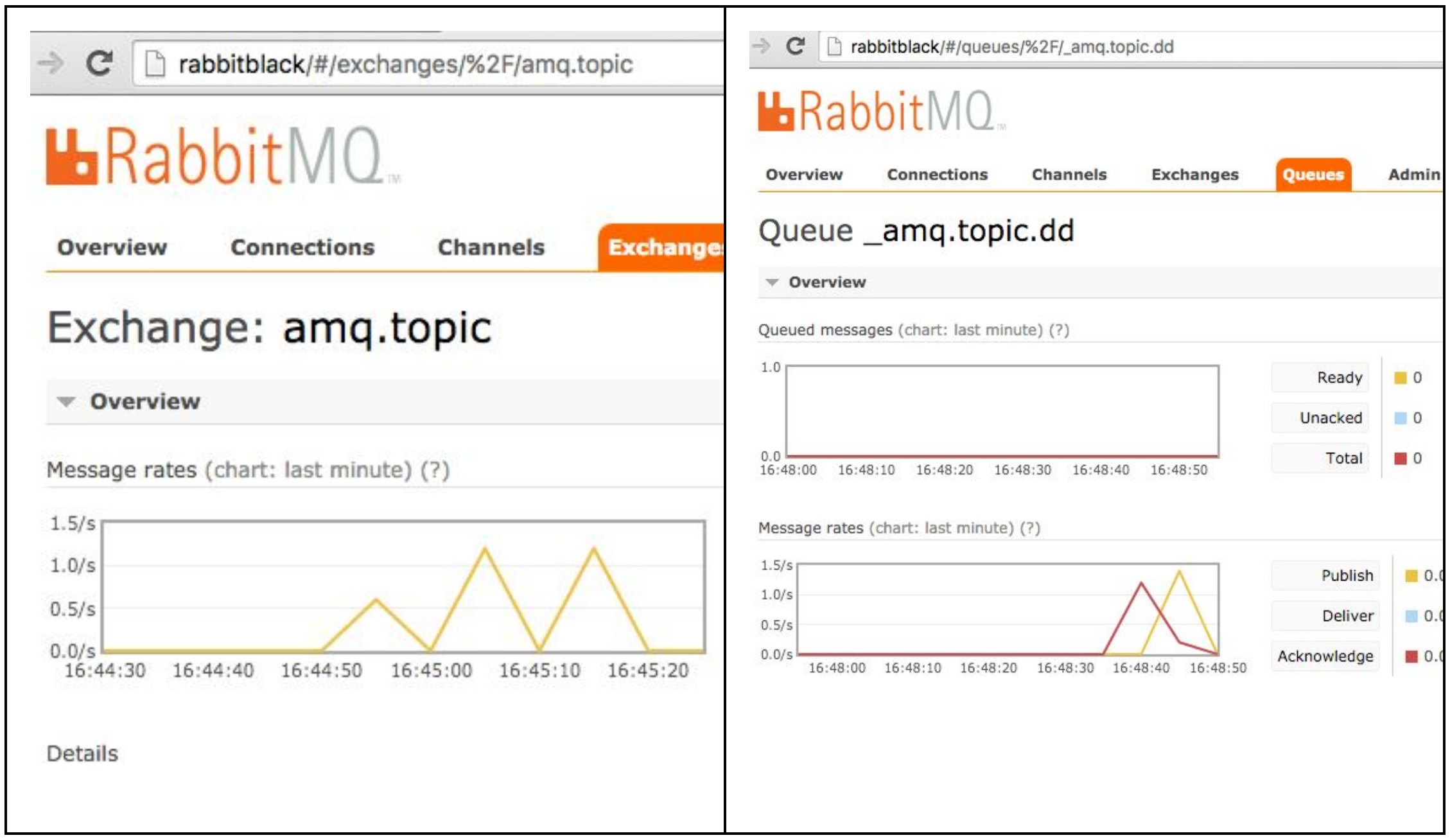

http://rabbitblack/#/exchanges/%2F/amq.topic http://rabbitred/#/exchanges/%2F/amq.topic

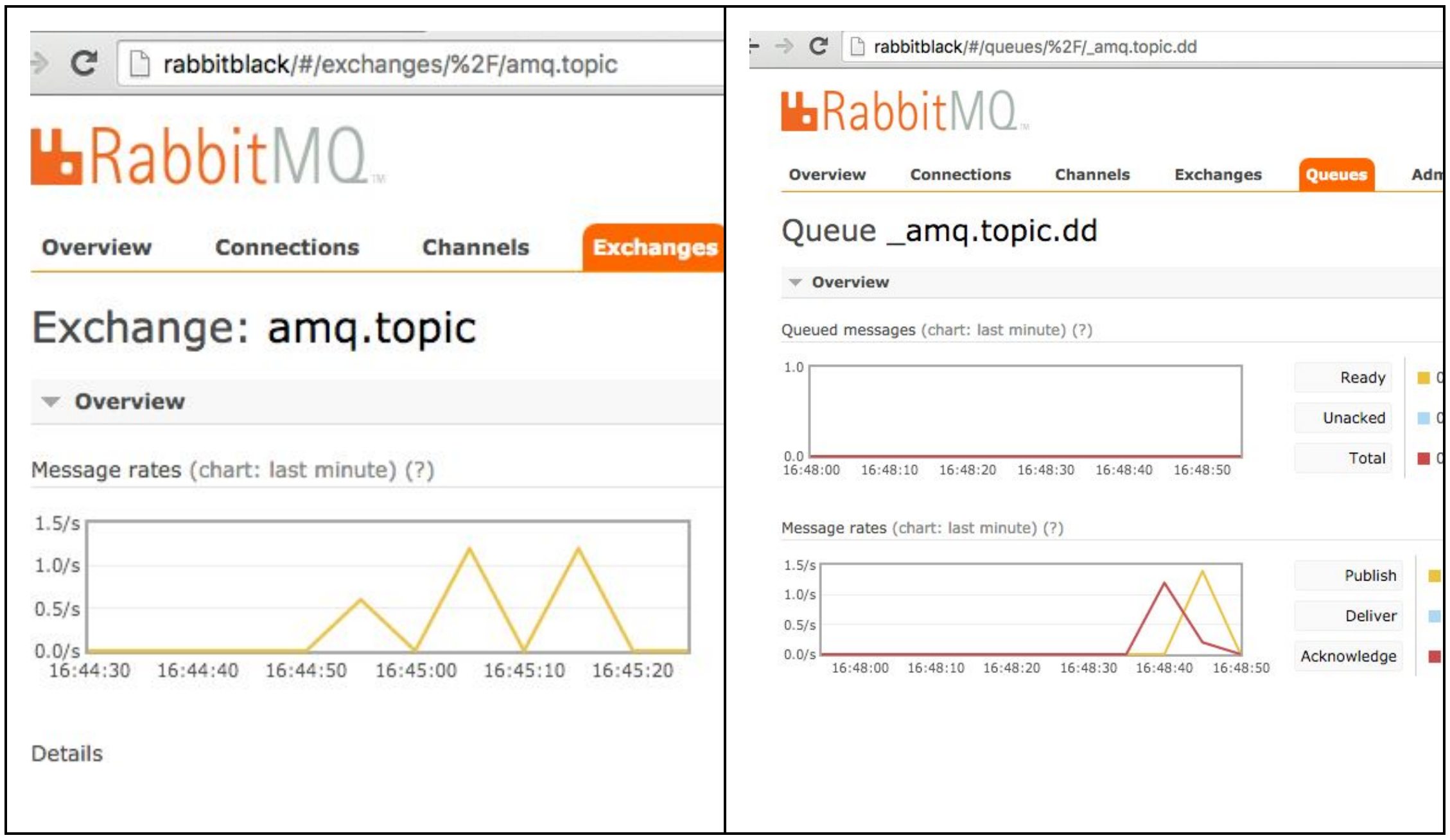

http://rabbitblack/#/exchanges/%2F/amq.topic        http://rabbitred/#/exchanges/%2F/amq.topic

The MQTT events will be send to the topic exchange but will not land in the queue on the red side because the messages contain a routing key. To get the MQTT messages from Node-RED, we bind the topic exchange with the routing key “timestamp”.

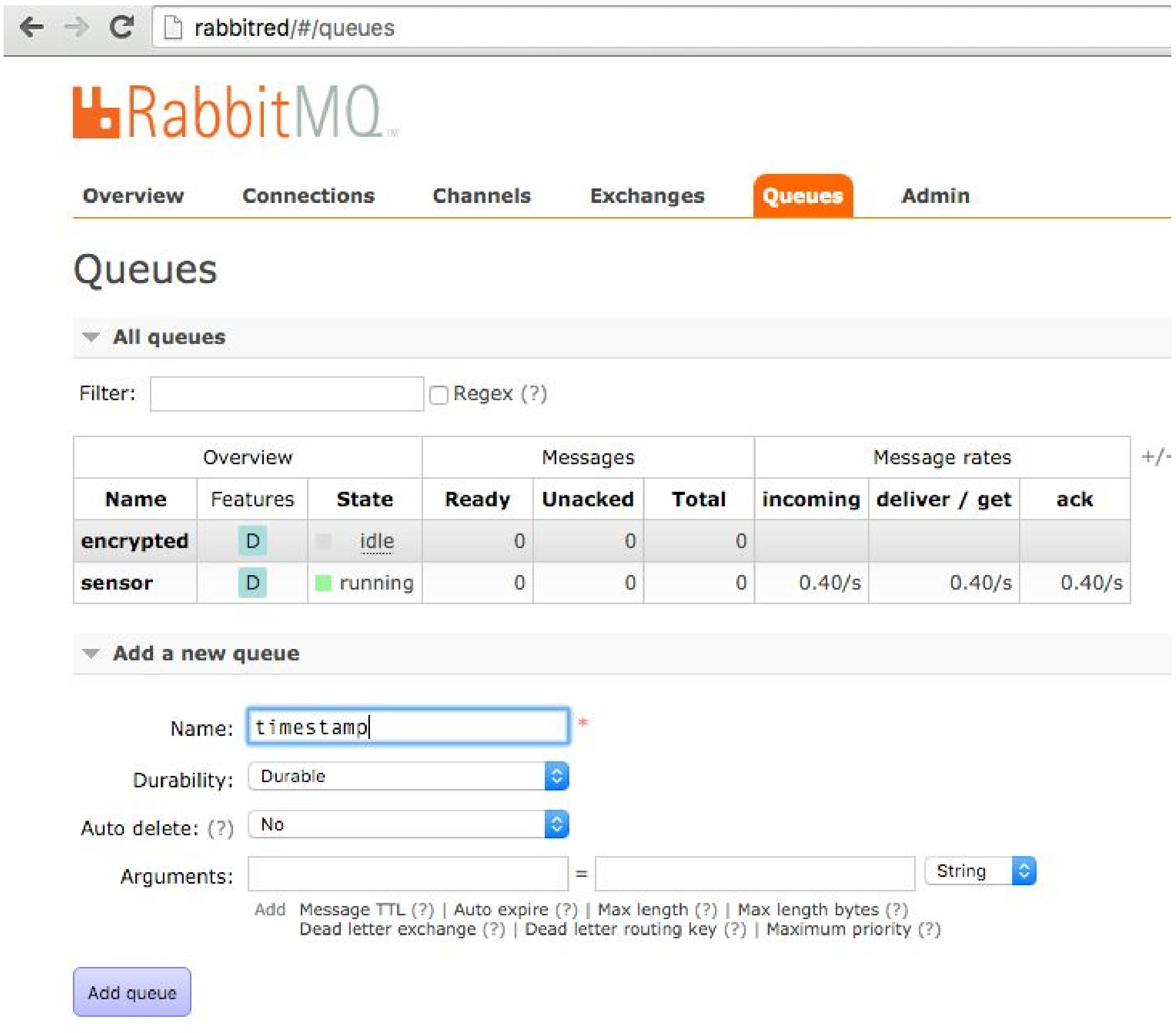

http://rabbitred/#/queues/

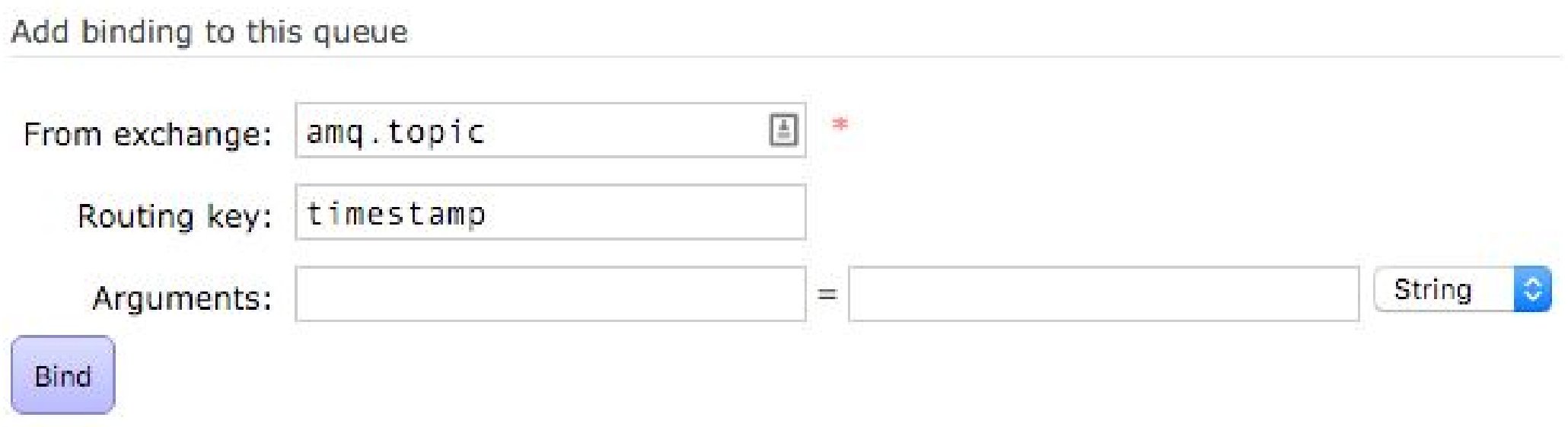

Now that the queue "timestamp" is bound to the exchange "amqp.topic" with the right routing key  messages are being collect at the queue on the inside.

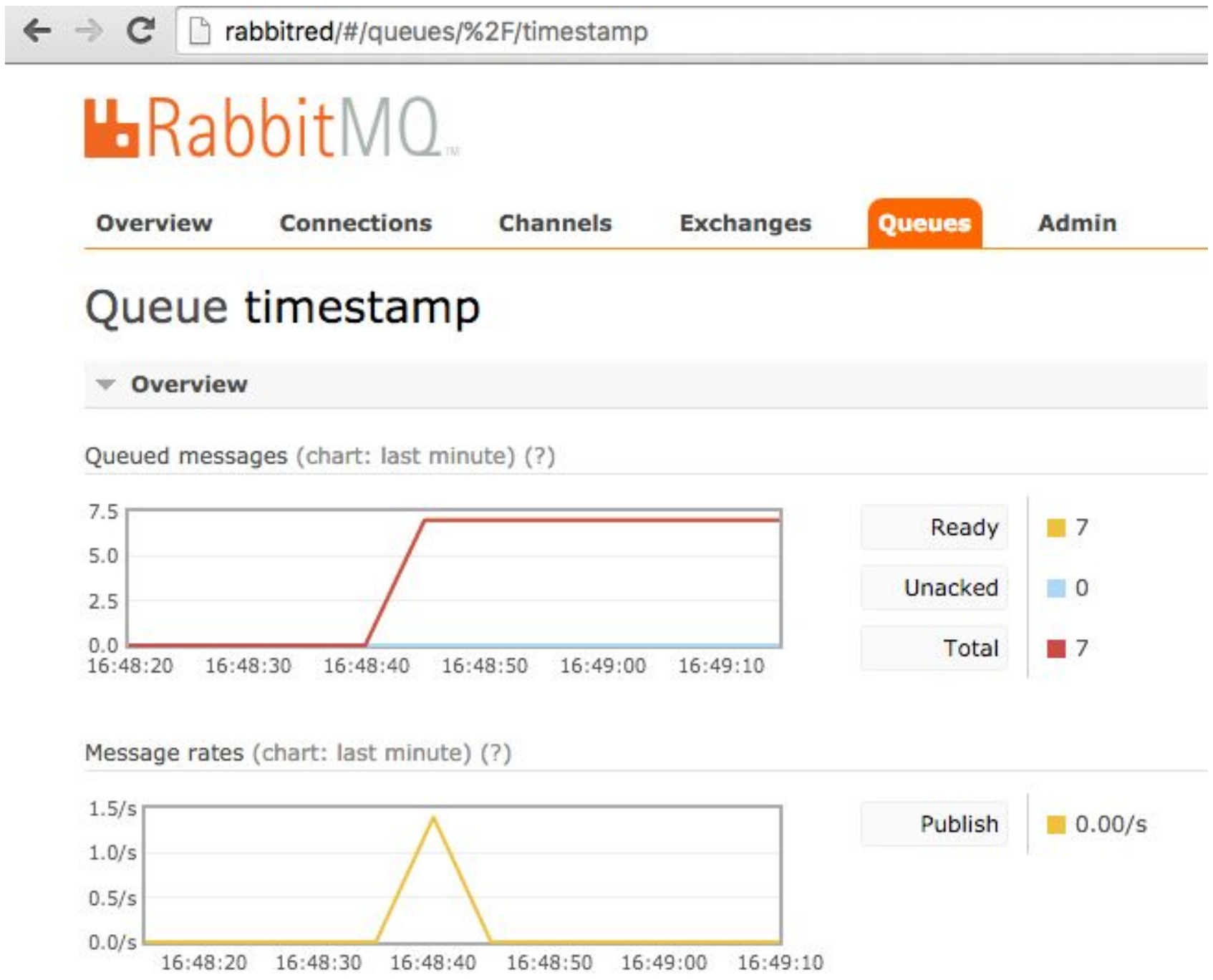

http://rabbitred/#/queues/%2F/timestamp

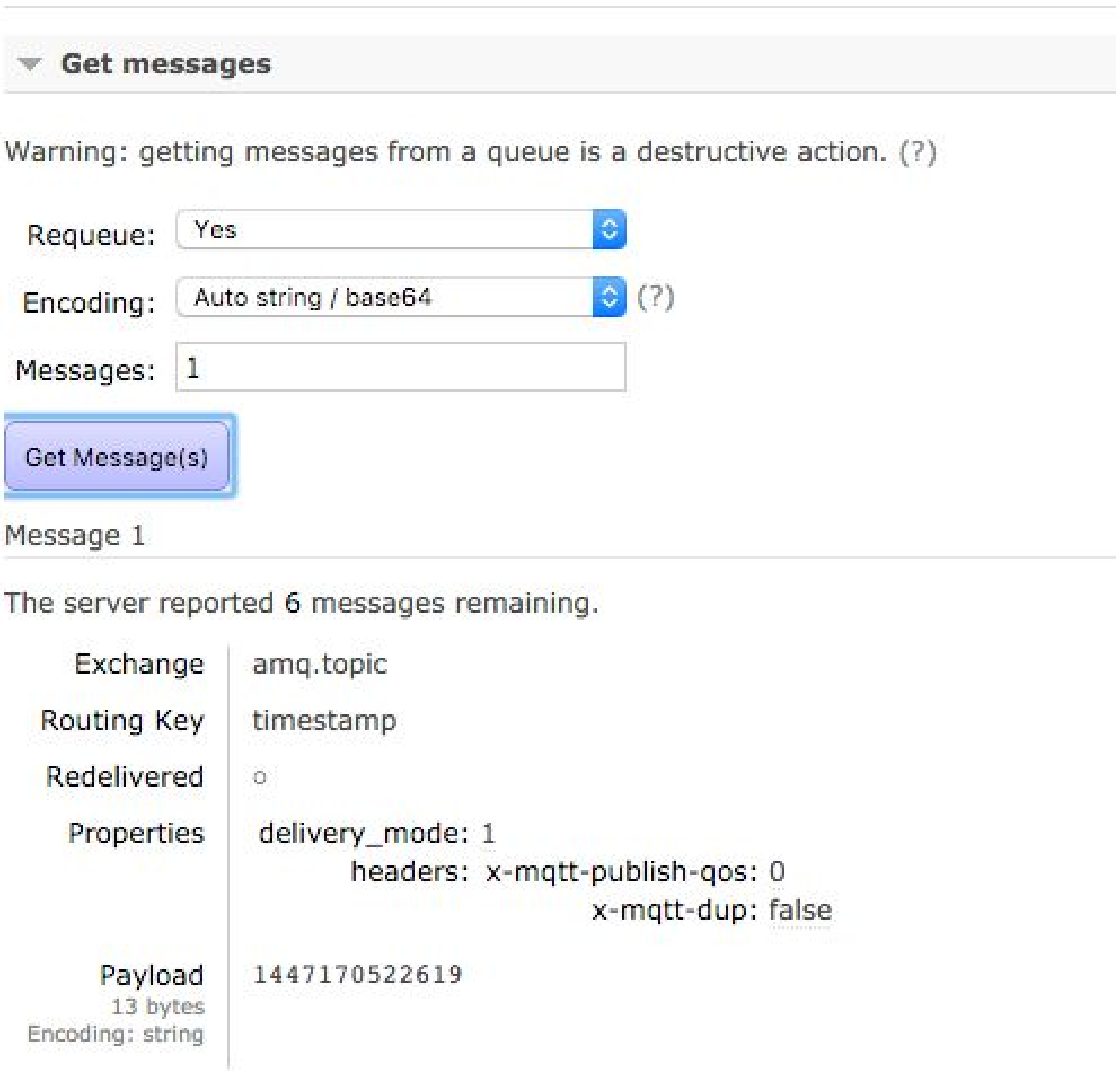

This is the contents one one of those MQTT messages.

**DOCKER-COMPOSE.YML**

```yaml
volumes:
 image: tianon/true
 volumes:
   - "./volumes/node-red/:/root/.node-red/"
   - "./volumes/rabbitmq/lib:/var/lib/rabbitmq/"
   - "./volumes/ldap/:/var/lib/ldap"
   - "./volumes/slap/:/etc/ldap/slapd.d"

proxy:
 image: jwilder/nginx-proxy
 ports:
   - "80:80"
volumes:
   - "/var/run/docker.sock:/tmp/docker.sock:ro"
 restart: always
 log_driver: "json-file"
 log_opt:
   max-size: "100k"
   max-file: "20"

rabbitblack:
 build: rabbitmq/
 restart: always
 environment:
   RABBITMQ_NODENAME: rabbitblack
   VIRTUAL_HOST: rabbitblack
   VIRTUAL_PORT: 15672
 hostname: rabbitblack
 volumes_from:
   - "volumes"
 ports:
   - "1884:1883"
   - "5673:5672"
   - "25673:25672"
   - "1234:1234/udp"
 expose:
   - "15672"
 links:
   - proxy
```

```yaml
rabbitred:
build: rabbitmq/
 restart: always
 environment:
   RABBITMQ_NODENAME: rabbitred
   VIRTUAL_HOST: rabbitred
   VIRTUAL_PORT: 15672
 hostname: rabbitred
 volumes_from:
   - "volumes"
 ports:
   - "1885:1883"
   - "5674:5672"
   - "25674:25672"
   - "1235:1234/udp"
 expose:
   - "15672"
 links:
   - proxy

nodered:
 image: beevelop/node-red
 environment:
   VIRTUAL_HOST: nodered
   VIRTUAL_PORT: 1880
 expose:
   - "1880"
 volumes_from:
   - "volumes"
 restart: always
 links:
   - rabbitblack
   - proxy
```

CONTENTS OF ENCRYPTED AND DECRYPTED CLASSES IN JSON

**SecureMessage.java:**

```
{
   "secureMessage":{
      "index":1,
       "signature":
              "EQE6lLamLrnf5QEaivpaW6XFUrZYQbLpuN0WI10qLQSX2MwLkMni9Iv4Kn
              oF/z9OIMPonk1EWNtn\nyza7JD+sPc1nyOd4frRs+NqWQqmJ7xGnFcnxAGv
              FjbVUg50kpCs2xu9f8F1GMLOhLHih3wKkiX5T\nfQTh/YgTjOQt/aeVTbd7
              W6JM479cB6D+9OQrhTQa/Yz+ut+0+ppBublX9drSizpoasnsWXWJbqLn\nC
              AG/xWow7co/zYcg9FznaOJXWaX/trDaKfL1WGmHIEXBNWHzlko8925uw3zL
              YGujmR5TlxHMidrz\n78l/AlasX3ur8yDizVRmuham1Tlj0PlXLclOHg==",
       "encryptedKey":
              "T6rqFFtYNxZ4C2GQFG7UGyzNH0tpvVNKlSB5jZZh5ATCF5H3NrHCNEsBU5
              LGlkomk0AsmSRH5JwQ\nFQQCSiEJa32OEroGFnqqKyK3csBkcchlPWsYwp6
              Z7dPmphMKK4gVzm5XGlGZ9La2m3JOqjLMQkAd\nxCXgxBGdhD9jyGl1XwGh
              mmsNlI1aOGB62F4lCz69IOqPjkKkRO6L6RakFZvdCId8glDoMxc3V/xu\nr
              Q090VfasifihKfaIwos4b/jaqEM+AJzYzaFKUX3turQ/JSatlt7TLTtJQKR
              xllbCpJOJkSHagxH\n99evUwDai5cBvjkZmVj+xea/mI9zaxKUTrmiVw==",
       "encryptedData":
              "MFKbRKjxCbcXAe+BdNt0nfsfhz0ARaJsnYzFCCoZWz4L/+/hdnuIakQCx0
              m4SBZRWE19xm1TaAsV\nh1qBz4WVXbayVP/OIhAqlscNtbceuESjmkHRqW8
              Bj334YsxPoT1nNy++A0MVmn7KpVFIhhvzdwB7\nqMJ9uJI9aMUwzj9RaizU
              GOmiMMuZfKo999+q36tNaVbhv4za7mk0LlgARRasXm0owSSvPnYLxIHw\n3
              ozh7q5bKHbBq4oENkfPcoOdqUHX4l+k3+tZxnD9Fk0ndRZTkCi4nGywfelM
              U4GM4wHewdeZlpiF\nRzA5VRXB3tSRIEc2whbiFECeB8+oLXI+JYCpE0Cln
              m2ITag1iOaA20XRDORAQhYHFPFaP/DlbzR7\nmP9hCUp4Yh9efg/PoM99Z0
              QhV/o7vmUscuzx7SABR/fnntvy/CFnfTWpNzS7TLyPxrbKaZ3IQbb\n6Vgv
              l2BUNf6Q/crgVtVslMfIjYNhHiAVUjtJmSjodXTtv2brW1KcIGAlbYBUxk9
              3P+w70wFwF8Hb\ndQ==",
       "iv":"ggAWY9MQYfcYs9PpZ6/kMg=="
   }
}
```

**ExchangeMessage.java:**

```
{
   "org.datadiode.model.message.ExchangeMessage":{
      "message":{
         "messageProperties":{
            "headers":[
               ""
```

```
            ],
            "contentType":"application/x-java-serialized-object",
            "contentLength":0,
            "contentLengthSet":false,
            "deliveryMode":"PERSISTENT",
            "priority":0,
            "redelivered":false,
            "receivedExchange":"sensor",
            "receivedRoutingKey":"bWFyY2Vs",
            "deliveryTag":1,
            "deliveryTagSet":true,
            "messageCount":0,
            "consumerTag":"amq.ctag-pwWjY2tzUjdS01cdVEnQQw",
            "consumerQueue":"encrypt"
         },
         "body":[

"rO0ABXNyAENvcmcuZGF0YWRpb2RlLm1vZGVsLmV2ZW50LnNlbnNvci50ZW1wZXJhdHVyZS5UZW1w\nZXJhdHVy
ZVNlbnNvckV2ZW50vqbEUzgjq5cCAAFEAAt0ZW1wZXJhdHVyZXhyACxvcmcuZGF0YWRp\nb2RlLm1vZGVsLmV2Z
W50LnNlbnNvci5TZW5zb3JFdmVudAmQNj9AL5ShAgABTAAGc2Vuc29ydAAp\nTG9yZy9kYXRhZGlvZGUvbW9kZW
wvZXZlbnQvc2Vuc29yL1NlbnNvcjt4cgAfb3JnLmRhdGFkaW9k\nZS5tb2RlbC5ldmVudC5FdmVudNLl636osM1
5AgADSQAFaW5kZXhMAARkYXRldAAQTGphdmEvdXRp\nbC9EYXRlO0wABHV1aWR0ABBMamF2YS91dGlsL1VVSUQ7
eHAAAAABc3IADmphdmEudXRpbC5EYXRl\naGqBAUtZdBkDAAB4cHcIAAABUMijD9J4c3IADmphdmEudXRpbC5VV
UlEvJkD95hthS8CAAJKAAxs\nZWFzdFNpZ0JpdHNKAAttb3N0U2lnQml0c3hwpkAZpTilJvLObegL2JVCfnNyAD
5vcmcuZGF0YWRp\nb2RlLm1vZGVsLmV2ZW50LnNlbnNvci50ZW1wZXJhdHVyZS5UZW1wZXJhdHVyZVNlbnNvcnj
ZPEmG\nuRW0AgAAeHIAJ29yZy5kYXRhZGlvZGUubW9kZWwuZXZlbnQuc2Vuc29yLlNlbnNvcjHcbb41TAJ6\nAg
AESQACaWRMAAtnZW9Mb2NhdGlvbnQAJ0xvcmcvZGF0YWRpb2RlL21vZGVsL2V2ZW50L0dlb0xv\nY2F0aW9uO0w
ACHRhcmdldGlkdAASTGphdmEvbGFuZy9TdHJpbmc7TAAEdHlwZXEAfgAOeHAAAAAB\nc3IAJW9yZy5kYXRhZGlv
ZGUubW9kZWwuZXZlbnQuR2VvTG9jYXRpb26jX4d5RdqQaAIAAkQACGxh\ndGl0dWRlRAAJbG9uZ2l0dWRleHBAS
jCJQ+EAYEATmQRwqAjIdAAIYldGeVkyVnN0AAt0ZW1wZXJh\ndHVyZUA1dmRNFdwU"
          ]
       },

"exchangeData":"{\"org.springframework.amqp.core.FanoutExchange\":{\"shouldDeclare\":tr
ue,\"declaringAdmins\":[{\"@class\":\"list\"}],\"name\":\"sensor\",\"durable\":true,\"a
utoDelete\":false,\"arguments\":[{\"@class\":\"linked-hash-map\"}]}}"
   }
}
```

**TemperatureSensorEvent.java**

```
{
   "org.datadiode.model.event.sensor.temperature.TemperatureSensorEvent":{
      "index":1,
      "date":"2015-11-02 14:39:01.74 UTC",
      "uuid":"ce6de80b-d895-427e-a640-19a538a526f2",
      "sensor":{
         "@class":"
           org.datadiode.model.event.sensor.temperature.TemperatureSensor",
         "type":"temperature",
         "id":1,
         "geoLocation":{
            "longitude":4.899431,
            "latitude":52.379189
         },
         "targetid":"bWFyY2Vs"
      },
      "temperature":21.4624679735535
   }
}
```